\documentclass[fleqn,usenatbib]{mnras}

\usepackage{newtxtext,newtxmath}

\usepackage[T1]{fontenc}

\DeclareRobustCommand{\VAN}[3]{#2}
\let\VANthebibliography\thebibliography
\def\thebibliography{\DeclareRobustCommand{\VAN}[3]{##3}\VANthebibliography}

\usepackage{graphicx}	
\usepackage{amsmath}	
\usepackage[version=4]{mhchem}

\defcitealias{venot_2019}{V19}

\title[Alkali Chemistry in Hot Jupiters]{Transport and Thermochemical Kinetics of Alkali Species in Hot Jupiter Atmospheres: Implications for Magnetic Models}

\author[D. A. Christie et al.]{
D. A. Christie,$^{1,2}$\thanks{E-mail: christie@mpia.de}
T. M. Evans-Soma,$^{3,1}$
N. J. Mayne,$^{2}$
E. H\'ebrard,$^2$
M. Zamyatina,$^2$,
J. E. Owen,$^{4,5}$\newauthor
and K. Kohary$^{2}$ 
\\
$^{1}$Max Planck Institute for Astronomy, K\"onigstuhl 17, D-69117 Heidelberg, Germany\\
$^{2}$Department of Physics and Astronomy, Faculty of Environment, Science and Economy, University of Exeter, Exeter EX4 4QL, UK\\
$^{3}$School of Science, University of Newcastle, Callaghan NSW, 2308, Australia \\
$^{4}$Imperial Astrophysics, Department of Physics, Imperial College London, Blackett Laboratory, Prince Consort Road, London SW7 2AZ, UK\\
$^{5}$Department of Earth, Planetary, and Space Sciences, University of California, Los Angeles, CA 90095, USA}
\date{Accepted 2026 September 22. Received 2026 August 28; in original form 2026 July 3}

\pubyear{\the\year{}}

\begin{document}
\label{firstpage}
\pagerange{\pageref{firstpage}--\pageref{lastpage}}
\maketitle

\begin{abstract}
Interactions between the planetary magnetic field of a hot Jupiter and the winds within its atmosphere are moderated by the electron and ion abundances. For a solar metallicity atmosphere at pressures between 10 and $10^{-4}$ bar, the primary sources for these charged species are expected to be the potassium and sodium atoms, and models of magnetic effects in hot Jupiter atmospheres have typically assumed that charged species exist within the atmosphere in their equilibrium abundances. In this work, we investigate the thermochemical kinetics of alkali species within a hot Jupiter atmosphere, thus allowing for the thermal ionisation fraction to depart from equilibrium, and demonstrate that quenching of ions occurs between 1 and 10 mbar. This increase in ionisation fraction on the nightside and at mid-latitudes results in stronger magnetic interaction, here modelled using the magnetic drag approximation, further slowing the jet and altering the global circulation. We additionally show that while quenching occurs, large magnetic resistivities resulting from low electron abundances continue to exist on the nightside between 10 and 100 mbar, in tension with magnetohydrodynamic models that assume horizontally uniform magnetic resistivities to simplify the explicit integration of the induction equation. At lower pressures ($\sim$ 0.1 mbar) where the electrons have become horizontally homogenised, the magnetic Reynolds number approaches or exceeds unity, signalling the breakdown of the magnetic drag approximation. We thus stress that chemical kinetics represents a dynamically important component of modelling hot Jupiter atmospheres while at the same time further complicating the treatment of magnetic effects.
\end{abstract}

\begin{keywords}
astrochemistry -- planets and satellites: gaseous planets -- planets and satellites: magnetic fields
\end{keywords}



\section{Introduction}
Hot and ultra-hot Jupiters are expected to have fast equatorial jets transporting heat from the dayside to the nightside \citep{showman_2008}, with dayside atmospheric temperatures sufficient to thermally-ionise alkali atoms \citep{perna_2010a,perna_2010b}. If the planet has a magnetic field, the charged species will experience Lorentz forces. Collisions involving the electrons and ions with hydrogen and helium will indirectly communicate these forces to the bulk neutral atmosphere. For sufficiently hot atmospheres and strong enough magnetic fields, it has been shown that this coupling can influence the atmospheric circulation, slowing the jet \citep{rauscher_2013,seidel_2026}, breaking equatorial symmetry \citep{batygin_2014,fecanin_2026}, and potentially suppressing the jet completely and inducing a day-night overturning flow \citep{beltz_2022}.

Modelling these magnetic effects in the atmospheres of hot Jupiters provides a number of challenges. In regions with sufficient ionisation, the equations of magnetohydrodynamics (MHD) can be solved and the evolution of the flow followed. This generally makes the problem of modelling the daysides of hot Jupiters tractable, if somehow the nightsides could be ignored. The nightsides of hot Jupiters, due to their relatively cool temperatures ($\sim 1000$ K compared to dayside temperature that can exceed 2000~K in some cases ) and lower thermal ionisation fractions at local thermochemical equilibrium, can be difficult to model as the magnetic field becomes largely decoupled from the atmosphere, requiring prohibitively small numerical time steps to solve the MHD equations explicitly\footnote{Within the single-fluid approach to non-ideal magnetohydrodynamics, the Ohmic term in the induction equation behaves like a diffusion operator, with a diffusion constant proportional to the resistivity $\eta$. The maximum explicit timestep based on the Courant-Friedrichs-Levy condition for numerical stability is thus proportional to $\eta^{-1}$. Given that $\eta \propto x_\mathrm{e}^{-1}$ \citep{perna_2010b} within the region of the atmosphere being considered, the low nightside ionisation requires small explicit time steps, typically orders of magnitude shorter than both the dynamical and radiative time steps adopted in simulations of hot Jupiters. The exact details, however, will depend on day-night heat redistribution and the nightside temperatures within the simulation.}. This can be mitigated by assuming an artificially large ionisation fraction on the nightside, as in \citet{rogers_2014} and \citet{rogers_2014b}, allowing the equations of non-ideal MHD to be solved on the nightside at the expense of the results not necessarily being accurate\footnote{\citet{rogers_2014} and \citet{rogers_2014b} also adopt horizontally uniform resistivities based on a reference pressure-temperature profile and thus do not necessarily capture the dayside changes in coupling.  This assumption was relaxed in \citet{rogers_2017}.}. An alternative approach is to abandon attempts to follow the evolution of the magnetic field and instead assume that any magnetic interactions can be reduced to a linear drag term applied to the atmosphere with a timescale depending on the local thermochemical conditions  (introduced for hot Jupiters in \citealt{perna_2010a,perna_2010b} and first included in a general circulation model (GCM) in \citealt{rauscher_2013}). Within this  ``magnetic drag''\footnote{ We do not include this class of models under the umbrella of {\em magnetohydrodynamic} models as the dynamics of the magnetic field is not included in the approach; however, this is not a universal distinction, and some authors refer to these methods as ``kinematic MHD'' \citep[e.g.,][]{beltz_2022}.} framework, the capturing of the decoupled field on the nightside becomes trivial as a decoupled atmosphere simply does not experience any forcing, and so long as the coupling on the dayside is relatively weak, any induced field should remain small and the simplifying assumption of a static dipole will be, at least approximately, valid. This approach has been used extensively to approximate magnetic effects in hot and ultra-hot Jupiters \citep[e.g.][]{beltz_2025,kennedy_2025,blocker_2026,frazier_2026} and has recently been adapted to investigate misaligned dipolar fields \citep{fecanin_2026}. Ultimately, these two approaches, each prioritising accuracy in one hemisphere at the expense of the other, are complementary.

Underlying the assertion that the nightside ionisation fraction is small is the assumption that the gas is in thermochemical equilibrium, with models either specifically focusing on the ionisation of sodium and potassium \citep[e.g.,][]{perna_2010a, perna_2010b,christie_2025} or attempting to account for the ionisation of all atomic species \citep[e.g.,][]{rauscher_2013,beltz_2022}. A more complete accounting of thermal equilibrium ionisation was done by \citet{blocker_2026} in their model of magnetic drag in the atmospheres of ultra-hot Jupiters, using pre-calculated tables generated using the thermochemical equilibrium chemistry code {\sc fastchem} \citep{kitzmann_2024}. Understanding the implications of this assumption raises two important questions. First, if the distribution of ions and electrons is allowed to deviate from thermochemical equilibrium and evolve via chemical kinetics, are the magnetic forces on the atmosphere altered? If departures from thermochemical equilibrium significantly change the magnetic forcing, the global circulation and the speed of the equatorial jet could be altered.  This could impact our understanding of transport-induced quenching\footnote{ Transport-induced quenching is the departure of abundances from chemical equilibrium due to transport processes such as advection or diffusion that results in their spatial homogenisation. } (the as in \citealt{zamyatina_2023}) and our ability to infer magnetic field strengths from wind speed observations (as in \citealt{seidel_2026}).  Second, as this assumption is central to the modelling difficulties associated with directly solving the equations of MHD, if the electrons and ions are not in thermochemical equilibrium, is it possible that the ionisation fraction is actually sufficiently high on the nightside to make the direct solution of the equations of non-ideal MHD tractable? If the answer to this second question is yes, it would support the results of \citet{rogers_2014} and \citet{rogers_2014b} where the magnetic field develops a large toroidal component, as those papers rely on an decreased nightside resistivity.

Thus, understanding the evolution and distribution of electrons and ions within the atmosphere is fundamental to understanding the forces on the atmosphere of these planets. In the 1000~K to 2000~K temperature range characteristic of hot Jupiters and for a solar elemental makeup, the primary sources of electrons are alkali species, predominantly potassium and sodium, due to their low ionisation energies (4.34~eV and 5.14~eV for potassium and sodium, respectively) and high elemental abundances relative to other species with similar ionisation energies. The existence of atomic alkali species in hot Jupiter atmospheres has been established observationally, first by the detection of the the sodium D line in transit observations of HD~209458~b \citep{charbonneau_2002}, with many other subsequent detections of sodium or potassium in the atmospheres of other planets (e.g., HD~189733~b, \citealt{redfield_2008}; WASP-6~b, \citealt{carter_2020}; WASP-39~b, \citealt{feinstein_2023}; WASP-69~b, \citealt{casasayas-barris_2017}; WASP-80~b, \citealt{sedaghati_2017}), although detections of their respective ions remain elusive.\footnote{Ions of other species with higher ionization energies have been observed in the atmospheres of ultra-hot Jupiters, such as \ce{Fe+} and \ce{Ca+} in MASCARA-2~b \citep{stangret_2020,nugroho_2020} and TOI-1518~b \citep{simonnin_2025}, and \ce{Fe+}, \ce{Ca+}, \ce{Ti+}, and \ce{Sc+} in KELT-9~b \citep{hoeijmakers_2019}. Some of these ionised species may trace escaping gas at lower pressures than we consider here. }  

Despite this observational evidence, modelling the chemical kinetics of neutral and ionised alkali species within a hot Jupiter atmosphere has been limited. \citet{koskinen_2010} modelled the ionisation of hydrogen atoms within the upper atmosphere of a HD~209458~b-like planet, between pressure of $10^{-5}$ bar and $10^{-10}$ bar, in a three-dimensional simulation and demonstrated that horizontal homogenisation of the ions due to horizontal winds is possible for hot Jupiters.  These lower pressures are characteristic of the regime in hot Jupiters where the atmosphere transitions to atomic hydrogen and EUV/XUV potentially regulate the launching of an escaping wind, not the deeper molecular layer around the infrared photosphere, and the modelling omitted heavier alkali species as a potential source of electrons.  Ions deeper in the atmosphere have been modelled by \citet{lavvas_2014} with a focus on potassium and sodium, formed by both photoionisation and thermal ionisation, in the atmosphere of HD~209458~b. The \citet{lavvas_2014} models are one dimensional, encompassing pressures between $10^3$ and $10^{-10}$ bar. Due to this limited dimensionality, they focus on the dayside atmosphere and only capture transport via diffusive mixing; however, they find photoionisation of potassium reaching as deep as 10 bar, while the ion abundances begin to deviate from thermochemical equilibrium at 0.1 bar. Although they make no claims about the nightside ionisation, they do demonstrate departures from thermochemical ionisation equilibrium are possible within the hot Jupiter parameter space, and thus these processes warrant further consideration.

In this paper, we model the thermochemical kinetics of alkali atoms in the atmospheres of hot Jupiters, with a focus on how the thermal ionisation of alkali atoms alters the distribution of electrons and ions and how that may impact models of magnetic drag. The structure of the paper is as follows: in Section \ref{Sec:Methods}, we introduce the 3D model and extended chemical network. In Section \ref{Sec:Results}, we discuss the results of the suite of simulations, highlighting the ability of quenching to increase the nightside ionisation. Finally, in Section \ref{Sec:Conclusions} we discuss the broader impacts for magnetic modelling and summarise the results. The paper also includes three appendices: the first comparing the chemical kinetics modelling to previous implementations (Appendix \ref{Appendix:V19Comp}), the second comparing three-body and radiative recombination (Appendix \ref{Appendix:RR}), and a final appendix containing additional figures (Appendix \ref{Appendix:AddPlots}).


\section{Methods}
\label{Sec:Methods}
Here we outline the modelling of the chemical kinetics of alkali atoms within the Met Office's {\sc unified model} (UM) three-dimensional GCM framework, including the effect of magnetic drag. We go into more detail on each aspect of the model below.

\subsection{The Unified Model}
The UM solves the non-hydrostatic, deep-atmosphere Euler equations on a spherical grid, and while it's provenance is in the modelling of the Earth, it has been adapted to model terrestrial and gaseous exoplanets \citep[e.g.,][]{mayne_2014a,amundsen_2016,mayne_2017}. The radiative transfer is solved using the {\sc socrates} code \citep{edwards_1996} which solves the radiative transfer in the two-stream, plane parallel limit and employs a pseudo-spherical approximation to compute the incoming stellar flux (\citealt{jackson_2020}; see also Appendix A of \citealt{christie_2022} for tests of the pseudo-spherical approximation). For the modelling of hot Jupiters, modifications to {\sc socrates} by \citet{amundsen_2014a,amundsen_2016,amundsen_2017} are used.

The opacity sources included are \ce{H2O}, \ce{CO}, \ce{CO2}, \ce{CH4}, \ce{HCN}, \ce{NH3}, \ce{K}, \ce{Na}, \ce{Li}, \ce{Cs}, and \ce{Rb}, as well as collision-induced absorption by \ce{H2-H2} and \ce{H2-He} (see Appendix A of \citealt{zamyatina_2023} for input opacity and broadening data). Except for \ce{Cs} and \ce{Rb}, the gas phase abundances for the relevant species are taken from the chemical solver (see Section \ref{Sec:Chem}), allowing for chemical disequilibrium to impact the gas opacity and thus the thermal structure, with the opacities mixed using equivalent extinction. As \ce{Cs} and \ce{Rb} are not in the chemical network used here, their gas-phase abundances are pre-calculated at thermochemical equilibrium and supplied to the radiative transfer solver using a look-up table.  

\subsubsection{Numerical Dissipation}

The model includes two explicit mechanisms of dissipation included to maintain numerical stability. The first is a `sponge' which dampens vertical motions near the upper boundary. Simulations performed here initially use a sponge in the upper quarter of the computational domain (in the terminology of \citealt{mayne_2014a}, the location of the start of the sponge, $\eta_\mathrm{s}$, measured as a fraction of the computational domain, would be $\eta_\mathrm{s}=0.75$) which is then raised to only encompass the upper 10\% of the computational domain ($\eta_\mathrm{s}=0.9$) over the first 300 days. As the initial state of the atmosphere in the simulation is spherically symmetric, simulations begin with nightside contraction due to cooling, and as a result, larger vertical motions. This approach effectively suppresses them while reducing the impact of the sponge for the majority of the simulation, as a deep sponge can limit vertical motions and artificially slow the jet in the upper atmosphere \citep{christie_2024}. The second is a longitudinal filter, effectively a diffusion operator with a grid-dependent diffusivity, which is applied to the horizontal velocities. This suppresses small-scale velocity fluctuations and atmosphere and maintains stability. We adopt a e-folding parameter $\Delta t_K=4$ to avoid the filtering being overly diffusive and shedding axial angular momentum. Details of the filter can be found in \citet{mayne_2014b,mayne_2014a} and \citet{christie_2024}.

\subsection{The Chemical Model}
\label{Sec:Chem}

To capture the departure of the ionisation state from thermochemical equilibrium, we extend the reduced thermochemical network of \citet[][hereafter V19]{venot_2019} with reactions relevant to alkali species as well as a number of charged species. The \citetalias{venot_2019} network provides C/N/O/H species relevant to the formation of the needed opacities and has been used in hot Jupiter models in, for example, \citet{zamyatina_2023,zamyatina_2024}. It lacks, however, the ionised species necessary to compute the gas conductivity which is required to model the magnetic interactions. We thus expand this network to include K-, Na-, and Li-bearing species motivated by \citet{lavvas_2014}. Each chemical species is stored as a tracer field within the GCM and is advected with the flow. The appropriateness of this assumption is discussed in Section \ref{Sec:AdvIons}.

\subsubsection{The Venot et al. (2019) Network}

The foundation of the chemical kinetics model is the C/N/O/H network of \citetalias{venot_2019} which is a reduced version the \citet{venot_2012} network and is sufficiently small in both number of reactions and species to be coupled directly to a GCM. It has been used with the UM's chemical kinetics solver to study chemical kinetics in hot Jupiter atmospheres \citep{drummond_2020, zamyatina_2023,zamyatina_2024} and has been benchmarked in \citet{christie_2026}. While more recent, updated C/N/O/H networks exist \citep[e.g.][]{veillet_2024}, they are too large to be realistically used in a coupled GCM simulation. The \citetalias{venot_2019} network also omits photochemistry which would require direct coupling to the radiative transfer routines, increasing the computational cost.

\subsubsection{Supplementary Chemistry}

\begin{table*}
\caption{Forward Reactions Supplementing the Venot et al. (2019) Reduced Network}
\label{Tbl:Chem}
\begin{tabular}{lcccc}
\hline
\hline
 & Reaction & Forward Rate (CGS)$^a$ & Source$^\mathrm{b}$ & Notes\\
\hline
\multicolumn{3}{l}{\em Sodium} \\
R1 & \ce{Na+ + e- + M <=> Na + M} & $k_0 = 3.43\times 10^{-14}T^{-3.77}$ & \citet{su_2001} & \\
   &                             & $k_\infty = 1\times 10^{-7}$ & & \\
R2 & \ce{NaH + H <=> Na + H2} & $k = 2.38\times 10^{-12}T^{0.69}\exp\left(-2360/T\right)$ & \citet{mayer_1966} & \\
R3 & \ce{Na + H2O <=> NaOH + H} & $k=4.1\times 10^{-10}\exp\left(-21900/T\right)$& \citet{jensen_1982} & \\
R4 & \ce{Na + OH +M <=> NaOH + M} & $k_0 = 1.9\times 10^{-25}T^{-2.21}\exp\left(-41/T\right)$ & \citet{patrick_1984} & \\
 & & $k_\infty = 1\times 10^{-11}$ & & \\
R5 & \ce{Na + H +M <=> NaH + M} & $k_0 = 1.9\times 10^{-25}T^{-2.21}\exp\left(-41/T\right)$ & & Estimated from R4$^\mathrm{c}$\\
 & & $k_\infty = 1\times 10^{-11}$ & & \\
R6 & \ce{Na+ + H- <=> Na + H} & $k=4.345\times 10^{-9}T^{-0.5}$ & & Estimate$^\mathrm{d}$ \\
R7 & \ce{Na + HCl <=> NaCl + H} & $k = 4\times 10^{-10}\exp\left(-4090/T\right)$ & \citet{husain_1986} & \\
\\
\multicolumn{3}{l}{\em Potassium} \\
R8 & \ce{K+ + e- + M <=> K + M} & $k_0 = 3.43\times 10^{-14}T^{-3.77}$ & & Estimate based on R1$^\mathrm{c}$ \\
 & & $k_\infty = 1\times 10^{-7}$ & & \\
R9 & \ce{KH + H <=> K + H2} & $k = 2.38\times 10^{-12}T^{0.69}\exp\left(-2360/T\right)$ & & Estimate based on R2$^\mathrm{c}$ \\
R10 & \ce{K + H2O <=> KOH + H} & $k=5\times 10^{-10}\exp\left(-20000/T\right)$& \citet{jensen_1979} & \\
R11 & \ce{K + OH + M <=> KOH + M} & $k_0 =2.66\times 10^{-25}T^{-2.21}\exp\left(-48/T\right) $ & \citet{patrick_1984} & \\
 & & $k_\infty = 1\times 10^{-11}$ & & \\
R12 & \ce{K + H + M <=> KH + M} & $k_0 =2.66\times 10^{-25}T^{-2.21}\exp\left(-48/T\right) $ & & Estimate based on R4$^\mathrm{c}$\\
 & & $k_\infty = 1\times 10^{-11}$ & & \\
R13 & \ce{K + HCl <=> KCl + H} & $k = 5.6\times 10^{-10}\exp\left(-4170/T\right)$ & \citet{husain_1988} & \\
R14 & \ce{K+ + H- <=> K + H} & $k=4.345\times 10^{-9}T^{-0.5}$ & & Estimate$^\mathrm{d}$ \\
\\
\multicolumn{3}{l}{\em Lithium} \\
R15 & \ce{Li^+ + e^- + M <=> Li + M} & $k_0 = 3.43\times 10^{-14}T^{-3.77}$ & & Estimate based on R1$^\mathrm{c}$ \\
 & & $k_\infty = 1\times 10^{-7}$ & & \\
R16 & \ce{Li + H2O <=> LiOH + H} & $k = 5.6\times 10^{-10}\exp\left(-7970/T\right)$ & \citet{plane_1998} & \\
R17 & \ce{Li + HCl <=> LiCl + H} & $k = 3.8\times 10^{-10}\exp\left(-883/T\right)$ & \citet{plane_1987} & \\
R18 & \ce{Li + OH + M <=> LiOH + M} & $k_0 = 1.83\times 10^{-25} T^{-2.19}\exp\left(-48/T\right)$ & \citet{patrick_1984} & \\
 & & $k_\infty = 1\times 10^{-11}$ & & \\
R19 & \ce{Li + H + M <=> LiH + M} & $k_0 = 1.83\times 10^{-25} T^{-2.19}\exp\left(-48/T\right)$ & & Estimate based on R18\\
 & & $k_\infty = 1\times 10^{-11}$ & & \\
R20 & \ce{LiH + H <=> Li + H2} & $1.593\times 10^{-12}T^{0.69}\exp\left(-2767/T\right)$ & \citet{mayer_1966} & \\
R21 & \ce{Li+ + H- <=> Li + H} & $k=4.345\times 10^{-9}T^{-0.5}$ & & Estimate$^\mathrm{d}$ \\
 \\
\multicolumn{3}{l}{\em Supplemental Chlorine Reactions} \\
R22 & \ce{H + HCl <=> H2 + Cl} & $k=2.4\times 10^{-11}\exp\left(-1730/T\right)$ & \citet{allison_1996} & \\
R23 & \ce{HCl + OH <=> H2O + Cl} & $k=6.84\times 10^{-19}T^{2.12}\exp\left(646/T\right)$ & \citet{bryukov_2006} \\
R24 & \ce{HCl + M <=> H + Cl + M} & $k_0=7.31\times 10^{-11}\exp\left(-41140/T\right)$ & \citet{baulch_1981} & \\
    & & $k_\infty = 1\times 10^{5}$ & & \\
\\
\multicolumn{3}{l}{\em Charge Exchange Between Sodium, Potassium, and Lithium} \\
R25 & \ce{Na+ + K <=> Na + K+} & $k=1\times 10^{-11}$ & & Estimate$^\mathrm{b}$\\
R26 & \ce{Li+ + K <=> Li + K+} & $k=1\times 10^{-11}$ & & Estimate$^\mathrm{b}$\\
R27 & \ce{Li+ + Na <=> Li + Na+} & $k=1\times 10^{-11}$ & & Estimate$^\mathrm{b}$\\
\\
\multicolumn{3}{l}{\em Miscellaneous Hydrogen Reaction} \\
R28 & \ce{H- + M <=> H + e- + M} & $k_0=6.74\times 10^{-17}T^2\exp\left(-19870/T\right)$ & \citet{huq_1982,huq_1983} & Fits by \citet{lenzuni_1991a} \\
 & & $k_\infty = 1\times 10^5$ & & \\
R29 & \ce{H- + H <=> H2 + e-} & $k = 4.671\times 10^{-10}T^{-0.39}\exp\left(-39.4 / T\right)$ & \citet{bruhns_2010} & Fit by \citet{millar_2024}\\
\hline
\multicolumn{5}{l}{$^\mathrm{a}$ {\footnotesize Low pressure and high pressure reaction rates are labelled as $k_0$ and $k_\infty$, respectively. Reaction rates without a pressure dependency are labelled as $k$. }}\\
\multicolumn{5}{l}{$^\mathrm{b}$ {\footnotesize To highlight limited rate availability, we only provide references for measured or theoretically-determined rates in this column. Sources for rates }}\\ 
\multicolumn{5}{l}{~~ {\footnotesize adopted by analogy or estimated as plausible values are instead discussed in the {\em Notes} column.}}\\
\multicolumn{5}{l}{$^\mathrm{c}$ {\footnotesize Estimate following \citet{lavvas_2014}. }} \\
\multicolumn{5}{l}{$^\mathrm{d}$ {\footnotesize Estimate from \citet{millar_2024}. }}
\end{tabular}
\end{table*}

To account for charged species, we extend the chemical network to include \ce{K+}, \ce{Na+}, and \ce{Li+} as well as molecules formed from \ce{K}, \ce{Na}, and \ce{Li}, with the additional reactions found in Table \ref{Tbl:Chem}. While alkali atoms \ce{Cs} and \ce{Rb} are also included as opacity sources, we do not include them within the extended chemical network due to a lack of available reaction rates, although we note that these species have relatively small abundances and are not expected to contribute significantly to the ionisation.

To include the relevant reactions for \ce{K} and \ce{Na} species, we follow \citet{lavvas_2014}, extended to include rates for \ce{Li} species. We note that there do exist a number of gaps in the available reaction rates, and \citet{lavvas_2014} fill in those rates using reasonable assumptions. For example, lacking reactions forming \ce{NaH}, the reaction rates for \ce{NaOH} are substituted (see reactions R4 and R5 in Table \ref{Tbl:Chem}). Additionally, high-pressure rates $k_\infty$ are unavailable for all reactions used here, and we follow \citet{lavvas_2014} in assuming the $k_\infty=10^{-11}\,\mathrm{cm^3s^{-1}}$ for three-body rates involving neutral species and $k_\infty=10^{-7}\,\mathrm{cm^3s^{-1}}$ for three-body rates involving ions. To combine the low pressure ($k_0$) and high pressure ($k_\infty$) rates for these supplementary reactions into the final rate $k$, we adopt the Lindemann form \citep{lindemann_1922},

\begin{equation}
k = k_\infty \left(\frac{P_\mathrm{r}}{1 + P_\mathrm{r}}\right),
\end{equation}

\noindent where $P_\mathrm{r} = \left[X\right]k_0 / k_\infty$ is the reduced pressure and $\left[X\right]$ is the concentration of the third body \ce{M} in the reaction. For the supplementary reactions we include here, we use the total gas number density for the third body concentration, and we do not apply any weighting to individual concentrations. The rates inherited from the \citet{venot_2019} network continue to use a subset of species in calculating the third body concentration and use varying weightings.

As the focus of this work requires an accounting of the ions and electrons, the most important reactions are the recombination of ions with electrons, and the reverse, the collisional ionisation of neutral atoms.  The forward rate for this reaction for \ce{Na} was measured by  \citet{su_2001} (see also Table \ref{Tbl:Chem}) and the reverse reaction can be computed through detailed balance and the relevant thermochemical data.  An earlier measurement of alkali three-body recombination rates by \citet{ashton_1973} found a lower recombination rate of $k=4.1\times 10^{-24}T^{-1}$ cm$^6$~s$^{-1}$, fit to measurements taken between $\sim 1900\,\mathrm{K}$ and $2600\,\mathrm{K}$, with the \citet{ashton_1973} measurements finding the three-body recombination rates to be insensitive to the alkali species in question.  The rate is a factor of three to seven lower than the \citet{su_2001} rate, with the two diverging at temperatures lower than 1900~K.  As the \citet{su_2001} rate is both more recent and is fit to measurements taken between 800~K and 2600~K, we adopt it as the recombination rate going forward. We do, however, take the insensitivity of the three-body recombination rate to the alkali species found by \citet{ashton_1973} as support for our adopting the \citet{su_2001} rate for the \ce{K} and \ce{Li} reactions as well.  This will result in similar quenching behaviours between the alkali ions.

The thermochemical data in the form of 7-term NASA Gibbs fits are taken from \citet{mcbride_1994} and are used to both compute chemical equilibria via Gibbs minimisation as well as reverse reaction rates. The exception is \ce{HCl} which is a 7-term fit that agrees with the updated 9-term fit in \citet{mcbride_2002}. The specific heat capacity calculated from this updated 7-term fit agrees with that of the 9-term fit to within 0.15\%.

Radiative recombination and photoionisation are not included in the model.  For pressures larger than $10^{-5}$ bar, the three-body recombination rate is greater than the radiative recombination rate (see Appendix \ref{Appendix:RR}), and thus the omission of the radiative recombination rate is not expected to alter the ionisation fraction significantly. Photoionisation of alkali species, conversely, could significantly increase the ionisation rate near the substellar point, with \citet{lavvas_2014} finding photoionisation of \ce{K} extending to $\sim 10$ bar in their model of HD~209458~b (see their Figure 4). We opt to omit photoionisation in this initial study and instead introduce it in a future work. As the EUV and XUV fluxes from the host star are often uncertain, this initial study of thermochemical ionisation will provide a lower limit on the quenching of charged species.

\subsubsection{Limitations of the Chemical Model}

Understanding the distribution of electrons and ions is essential for the calculation of the conductivity, and with it, the magnetic forcing on the atmosphere. While this work addresses the importance of chemical kinetics, it remains that the chemical model has a number of aspects that could be improved upon.

First and foremost, the chemical network here represents a best attempt based on available reaction rates. As discussed above, in many cases, reaction rates do not exist and comparable reactions involving different species have been used in their places. Furthermore, unlike C/N/O/H networks \citep[e.g.,][]{venot_2012,veillet_2024}, it has not been validated as a whole against laboratory experiments. As chemical networks used in the modelling of exoplanetary atmospheres expand to account for additional species being observed, the need for laboratory measurements of reaction rates and network validation will only become even more important.

Second, the recombination of ions only includes recombination directly to the ground state without any accounting for thermally excited states. While rates and cross-sections for these reactions are available for ground states of the atoms in question \citep{verner_1996,verner_1996a}, they are not for all of their excited states. Modelling of the lowest excited states of \ce{K} and \ce{Na} were done by \citet{lavvas_2014}; however, adopting this model would increase the number of species in the network, increasing the computational cost, and has thus been omitted. Excited alkali atoms could, potentially, provide a lower energy pathway to thermal and photoionisation. Similarly, ignoring recombination to excited states results in an underestimation of the total recombination rate. 

Third, we have omitted photoionisation from this investigation. This would be expected to enhance the ionisation rate around the substellar point \citep[see][]{lavvas_2014} with the possibility of increasing the quenched abundance of electrons on the nightside. The impact of any future inclusion of photoionisation will depend on the stellar EUV/XUV flux as well as the degree of attenuation due to the atmosphere above the modelled computational domain, adding additional uncertain parameters. In terms of magnetic models, the inclusion of photoionisation would be expected to increase the electron and ion abundances around the substellar point, and thus increasing drag, especially on the dayside, and have the potential to further alter the global circulation. As potassium  has the lowest ionization energy of potassium, lithium, and sodium, photons capable of ionising potassium are less likely to be attenuated and thus be the main source of differences between a the model introduced here and a photochemical model. This is contingent on NUV photons reaching to pressures investigated here ($\sim 1$ mbar).  Clouds and hazes may provide an additional source of NUV absorption, reducing any photoionisation of potassium. 

We finally note that elemental abundances of alkali species in hot Jupiter atmospheres are not always well constrained. While photoionisation may hinder observation of, for example, sodium or potassium \citep{fortney_2003}, it may also be the case that alkali elements are simply depleted relative to other elements within the atmosphere, altering the chemistry and ionisation fraction. What we present here is a theoretical investigation of the effects of chemical kinetics on magnetic drag models, and the intention is not to compare the results to observations of any particular planet; however, in cases where the intent is to match observational results, this should be viewed as an additional source of uncertainty.

\subsection{Magnetic Drag Model}

We adopt the magnetic drag model introduced in \citet{christie_2025} which approximates all magnetic effects on the atmosphere by a drag on the gas proportional to the velocity component perpendicular to the local magnetic field,

\begin{equation}
\left(\frac{\partial \mathbf{u}}{\partial t}\right)_\mathrm{magnetic} = -\frac{\mathbf{u}_\perp}{\tau_\mathrm{drag}}\,\, ,
\label{Eqn:Drag}
\end{equation}

\noindent where $\mathbf{u} = (u,v,w)$ is the local velocity and $\mathbf{u}_\perp=-\left(\mathbf{u\times b}\right)\times\mathbf{b}$ is the component of the velocity perpendicular to the local magnetic field. The direction of the local magnetic field is given by the unit vector $\mathbf{b}$, and the drag timescale $\tau_\mathrm{drag}$ is 

\begin{equation}
\tau_\mathrm{drag} = \frac{c^2\rho}{\sigma_\perp B^2},
\end{equation}

\noindent where $\rho$ is the gas density, $B$ is the local magnetic field strength, and $\sigma_\perp$ is the perpendicular conductivity. The conductivity is computed as in \citet{christie_2025} and captures the effect of both Ohmic dissipation and ambipolar diffusion on the drag timescale. This also builds upon the model of \citet{perna_2010a} and \citet{perna_2010b} by accounting for drag in the vertical and meridional directions, and allowing for drag at the magnetic equator, which is not included in the prior models (see \citealt{christie_2025} for a more complete discussion). The geometry of the magnetic field is assumed to be dipolar, aligned with the rotational axis of the planet.

\begin{align}
        B_{\lambda} & = 0 \nonumber\,\, ,\\
        B_{\theta} & = \frac{B_\mathrm{ref}}{2}\left(\frac{r_\mathrm{ref}}{r}\right)^3\cos\theta \,\, ,\nonumber\\
        B_{r} & = B_\mathrm{ref}\left(\frac{r_\mathrm{ref}}{r}\right)^3\sin\theta\,\, ,
        \label{Eqn:B}
\end{align}

\noindent where $\lambda$, $\theta$, and $r$ are the longitude, latitude, and radius, respectively, and $B_\mathrm{ref}$ (with an adopted value of $10$~G) is the magnetic field strength at a prescribed radius $r_\mathrm{ref}$ (with an adopted value of $9.5\times 10^7$~m, roughly the middle of the computational domain) along the polar axis . This results in the magnetic field strength varying both radially and latitudinally, with the field at the same radii stronger near the poles relative to the equator. This leads to the local field strength on an isobaric surface varying not only because of the latitudinal variations but also due to the fact that isobaric surfaces do not correspond to surfaces of constant radius.  Adopting an aligned dipole is a simplifying assumption as the tilt of  any hot Jupiter magnetic field relative to its rotational axis is unconstrained by observation.  Within the solar system, Jupiter has a dipolar field tilted by roughly 10\textdegree \citep{connerney_2017,connerney_2018}, while Saturn has a tilt of less than 1\textdegree \citep{dougherty_2018}. Thus, while there is no evidence that the alignment between the hot Jupiter magnetic dipole  and rotational axis should be perfect, the assumption that it is is not unreasonable.  This uncertainty does motivate further investigation of how a tilted dipole may alter the circulation. This has been investigated analytically by \citet{batygin_2014} and using GCM simulations of ultra-hot Jupiters by \citet{fecanin_2026}, with each work demonstrating that the field is capable of breaking the north-south symmetry of the jet. The conclusions of these authors is unlikely to be changed by the inclusion of chemical kinetics, but any investigation of how a misaligned dipole alters our own conclusions is beyond the scope of this paper.

In Equation \ref{Eqn:Drag}, we omit the Hall term, as it contributes negligibly at the temperatures investigated here (see \citealt{christie_2025} for a discussion of the impact in hot Jupiters). At the hotter temperatures found on the dayside of ultra-hot Jupiters, however, it can result in a deflection of the flow, as shown by \citet{blocker_2026} in their investigation of WASP-18~b, and we refer the reader to their investigation for a thorough discussion of the impact of that term.

\subsubsection{Advection of Ions}
\label{Sec:AdvIons}

We do not advect the ions and electrons separately from the bulk neutral atmosphere. While separation between the charged and neutral species may occur as the charged species become attached to the magnetic field, for the regions of interest examined here we remain in the Ohmic limit where all charged species move in tandem with the neutrals. This occurs when the magnetisation $M_s$ of species $s$ satisfies $M_s \equiv \omega_s\tau_{s\mathrm{n}} \ll 1$ where $\omega_s=qB/mc$ is the gyrofrequency and $\tau_{s\mathrm{n}}$ is the momentum exchange timescale for a particle of species $s$ with charge $q$ in a gas of neutrals. This assumption breaks down at low pressures, where collisions between charged and neutral species are infrequent enough to allow the charged species to couple to the field. For the simulations presented here, the lowest pressure isobaric surface entirely within the computational domain at a pressure of $P=3\times 10^{-5}$ bar\footnote{On the night side, where the atmosphere is cooler, the pressures reach $\sim 10^{-8}$ bar; however, these regions will be significantly impacted by the upper boundary as the isobaric surfaces do not extend to the dayside.}, and on this surface $\omega_\mathrm{K}\tau_\mathrm{Kn} \sim 0.044$ to $0.121$. The variability across the isobaric surface is primarily due to the spatial variation of the magnetic field strength resulting from the dipole field geometry, with the field strength along the isobaric surface reaching 4.27 G near the equator and 8.91 G at the pole. The simplifying assumption that the ions move with the flow is thus reasonable as this estimate is at near the upper boundary, and deeper in the atmosphere the collisional coupling between the ions and neutrals will only be increased.  Relative to the ions, however, the electrons begin to decouple from the neutral atmosphere deeper in the atmosphere ($\sim 10^{-3}$ bar), causing $\sigma_\perp$ to deviate from the Ohmic limit. This results in an increase in the magnetic drag timescale at low pressures relative purely Ohmic models of magnetic drag (see Figures 2 and 3 in \citealt{christie_2025}).

\subsubsection{Limitations of Magnetic Drag Models}
\label{Sec:LimitMag}
Magnetic drag models allow for the influence of magnetic effects to be approximated and are especially useful for planets where the nightside ionisation is sufficiently low that the magnetic field is largely decoupled from the atmosphere, a regime that poses challenges for methods that explicitly solve the induction equation. This approach does come with some limitations, and it is important not to over-interpret results.

As the background magnetic field is static, there is no possibility for the magnetic field to adapt to the flow. Simulations that have solved for the evolution of the magnetic field directly, such as \citet{rogers_2014} and \citet{rogers_2014b}, find that the initially dipolar magnetic field is able to wind around the equator, increasing the magnetic field strength and significantly altering the field morphology (see Figure 3 in \citealt{rogers_2014b} for an illustration of the field deformation). While these studies do adopt artificially large ionisation fractions on the nightside to make the problem tractable, and there do exist a number of other caveats when interpreting their results (see the discussion in \citealt{christie_2025}), the fact that they find that the dayside field deforms significantly should motivate caution.

Often, the magnetic Reynolds number is invoked to justify the use of the magnetic drag approximation. The inclusion of drag replaces the solution of the magnetic induction equation\footnote{For the sake of simplicity, we omit the ambipolar and Hall terms in the induction equation for the purposes of this discussion.},

\begin{equation}
\frac{\partial \mathbf{B}}{\partial t} = \mathbf{\nabla}\times\left(\mathbf{u}\times \mathbf{B}\right) - \frac{c^2}{4\pi}\mathbf{\nabla}\times\left(\eta\mathbf{\nabla}\times\mathbf{B}\right)\,
\label{Eqn:Induction}
\end{equation}

\noindent with the second term accounting for the advection of the magnetic field and the third term the diffusion of the field through the neutral gas. The ratio of these two terms, along with the assumption that $\nabla\rightarrow L^{-1}$ and $\mathbf{u} \rightarrow V$ yields the magnetic Reynolds number $R_\mathrm{m}$,
\begin{equation}
R_\mathrm{m} = \frac{4\pi L V}{c^2\eta},
\label{Eqn:Rm}
\end{equation}

\noindent for a characteristic length scale $L$, velocity $V$, and Ohmic resistivity. If $R_\mathrm{m} \ll 1$, the advection of the field is presumed to be negligible and the magnetic drag prescription can be applied. If $R_\mathrm{m} \gg 1$, the field is presumed to be advected with the flow resulting in a deformation of the field, and the magnetic drag approximation likely breaks down. The validity of this interpretation depends on the choice of $L$ and $V$. The standard practice has been to evaluate $R_\mathrm{m}$ with either zonal or horizontal wind speeds from simulations including magnetic drag \citep[e.g.,][]{rauscher_2013,beltz_2022,christie_2025,blocker_2026}. This has the potential for introducing undue confidence in the results, as it assumes that the magnetic drag prescription (Equation \ref{Eqn:Drag}) reasonably approximates the Lorentz forces derived when solving the induction equation (Equation \ref{Eqn:Induction}). If the solution to the magnetic drag equations begins to deviate from the solution to the MHD equations, it is not necessarily the case that the magnetic Reynolds number resulting from the simulated atmosphere will accurately inform us of the relative importance of the advection and diffusion terms. Thus, regimes that significantly alter the flow, such as the {\em magnetic regime} in \citet{beltz_2022} wherein the zonal jet is suppressed and the flow is instead over the poles or the substellar magnetic {\em dead zone} in \citet{christie_2025}, may be unphysical despite having $R_\mathrm{m}\sim 1$. There is no opportunity for the field to deform and accommodate the flow, and the natural result of high ionisation in the magnetic drag formalism is a suppressed flow, which will reduce $V$ and thus $R_\mathrm{m}$. 

Out of caution, when analysing the magnetic Reynolds number, we analyse simulations with and without magnetic drag applied, and do not extend our suite of simulations to hotter cases where the flow is suppressed.

\subsection{Suite of Simulations}

As in \citet{christie_2025}, we adopt parameters of a hot Jupiter based on HD~209458~b, as it represents a standard test case for theoretical models and a hot Jupiter where magnetic drag is not expected to drastically reshape the flow. The common parameters for all simulations are shown in Table \ref{Tbl:Common}. In addition to the fiducial irradiation case, we adopt an enhanced radiation case, with an instellation $1.5$ times larger than HD~209458~b, in order to investigate hotter atmospheres with higher ionisation fractions while keeping the remaining parameters fixed. We assume a solar elemental composition \citep{caffau_2011, asplund_2009} with a C/O ratio of 0.55 for the ease of comparison with previous models; however, we note that this may not reflect the actual elemental composition of HD~209458~b (e.g., \citealt{xue_2024} find a metallicity of $3\times$ solar with a C/O ratio of 0.11 while \citealt{bachmann_2025} find a metallicity of $\sim 1.3\times$ solar with a C/O ratio of 0.054).

For each instellation case, we perform four simulations: the first investigating chemical equilibrium and chemical kinetics cases to serve as a baseline and second, the same simulations except with magnetic drag included. The simulations including magnetic drag adopt a reference pressure of $B_\mathrm{ref} = 10\,\mathrm{G}$ as this is sufficiently strong for magnetic drag to impact the jet, as demonstrated in \citet{christie_2025}. Due to the spatial variation in the field strength within the computational domain (see Eqn. \ref{Eqn:B}), the local field strength will vary between the polar value of 11.76 G at the inner boundary and the equatorial value of 4.286 G at the outer boundary. As the conductivity and thus the drag timescale depends on the local field strength, this can reduce the impact at lower pressures. While other investigations of magnetic drag take their fiducial field strength to be 3~G \citep[e.g.,][]{rauscher_2013}, their adopted field strength is constant in space, and so the larger reference value at the pole here does corresponds to a lower value by roughly a factor of two at the equator where the magnetic effects are expected to be most important. 

Additionally, to understand how the extended network compares to the more standard approach of employing chemical kinetics only for the V19 network while including alkali species through lookup tables, we perform two simulations that separate the calculations of the C/N/O/H and alkali species. These tests are discussed in Appendix \ref{Appendix:V19Comp}.

\begin{table}
\caption{Common Simulation Parameters}
\label{Tbl:Common}
\begin{tabular}{lcc}
\hline
\hline
  & Value & Units\\
\hline
{\em Grid and Time-stepping} \\
Longitude Cells & 144 &\\
Latitude Cells & 90 &\\
Vertical Layers & 80 & \\
Hydrodynamic Timestep & 30 & s \\
Chemistry Timestep & 3750 & s \\
Simulation Length & 1000 & Earth days \\
\\
{\em Radiative Transfer} \\
Bands & 32 & \\
Wavelength Range & 0.2 - 200 & $\mathrm{\mu}$m \\
$F_{\star,\mathrm{HD209458b}}$ (at 1 AU) & 2054.73 & $\mathrm{W\, m^{-2}}$ \\
Radiation Timestep & 150 & s \\
\\
{\em Damping and Diffusion} \\
Damping Profile & Horizontal &\\
Damping Coefficient & 0.15 &\\
Damping Depth ($\eta_s$) & 0.9 & \\
Filter Parameter ($t_K$) & 4 & \\
\\
{\em Planet}\\
Intrinsic Temperature ($T_\mathrm{int}$) & 100 & K \\
Initial Inner Boundary Pressure & 200 & bar \\
Inner Boundary Radius ($R_\mathrm{inner}$) & $9.0\times 10^7$ & m \\
Gravitational acceleration at $R_\mathrm{inner}$ ($g$) & 10.79 & $\mathrm{m\,s^{-2}}$\\
Semi-major axis ($a$) & $4.747\times 10^{-2}$ & AU \\
Ideal Gas Constant ($R$) & 3556.8 & $\mathrm{J\,kg^{-1}\,K^{-1}}$ \\
Specific Heat Capacity ($c_\mathrm{P}$) & $1.3\times 10^4$ & $\mathrm{J\,kg^{-1}\,K^{-1}}$ \\
Angular velocity ($\Omega$) & $2.06\times 10^{-5}$ & $\mathrm{s^{-1}}$ \\
Magnetic Reference Radius ($r_\mathrm{ref}$) & $9.5\times 10^{7}$ & m \\
Reference Field Strength ($B_\mathrm{ref}$) & $10$ & G \\
\hline
\end{tabular}
\end{table}

\section{Results}
\label{Sec:Results}

We now present the results from the suite of simulations outlined above. We adopt the UM's spherical coordinate system, with a longitude of $\lambda=0^\circ$ and latitude of $\theta=0^\circ$ corresponding to the anti-stellar point and a longitude of $\lambda=180^\circ$ and latitude of $\theta=0^\circ$ corresponding to the substellar point. The morning and evening terminators are at $\lambda=90^\circ$ and $\lambda=270^\circ$, respectively.

\subsection{Thermal Structure}
\label{Sec:Thermal}

\begin{figure*}
	\includegraphics[]{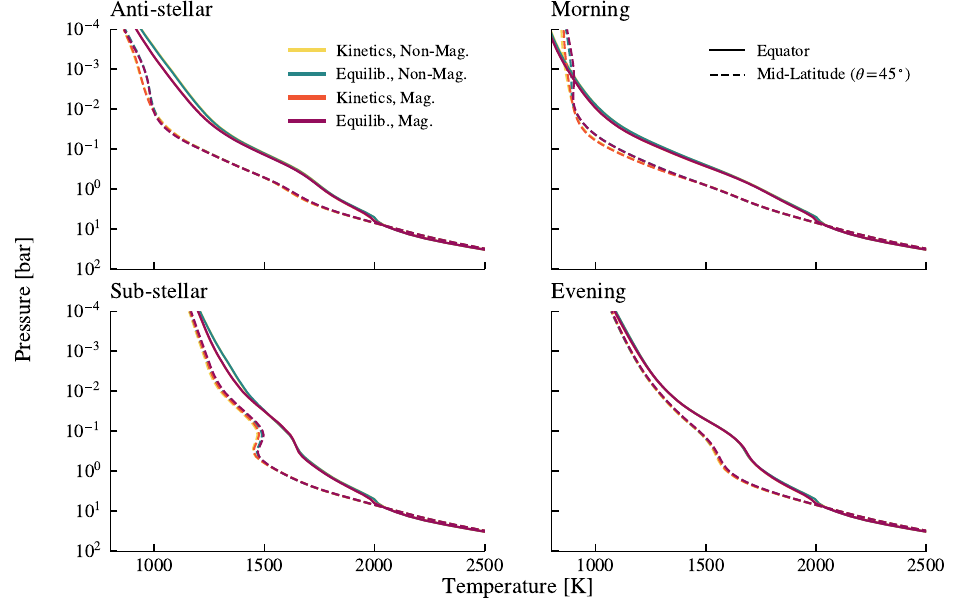}
    \caption{Equatorial (solid lines) and mid-latitude (dashed lines) pressure-temperature profiles for each of the simulations with the fiducial instellation.   Note that the equilibrium and chemical kinetics plots overlap and are thus difficult to distinguish. The largest discernable differences are due to the differences between magnetic and non-magnetic simulations.} 
    \label{Fig:pt_1x}
\end{figure*}

\begin{figure*}
	\includegraphics[]{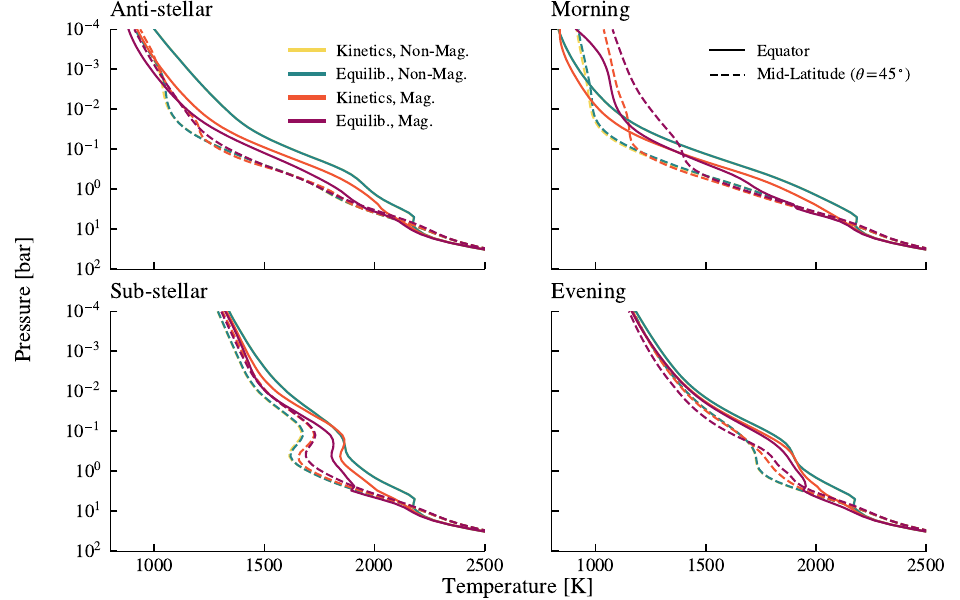}
    \caption{Equatorial (solid lines) and mid-latitude (dashed lines) pressure-temperature profiles for each of the simulations with the $1.5\times$ the fiducial instellation. Larger temperature differences between the simulations are observed here compared to in the fiducial case, up to 225 K between the magnetic and non-magnetic models and up to 170 K between the magnetic models.  See the text for further details.} 
    \label{Fig:pt_15x}
\end{figure*}

\begin{figure*}
	\includegraphics[alt={Four panels each with temperature contours for each simulation.}]{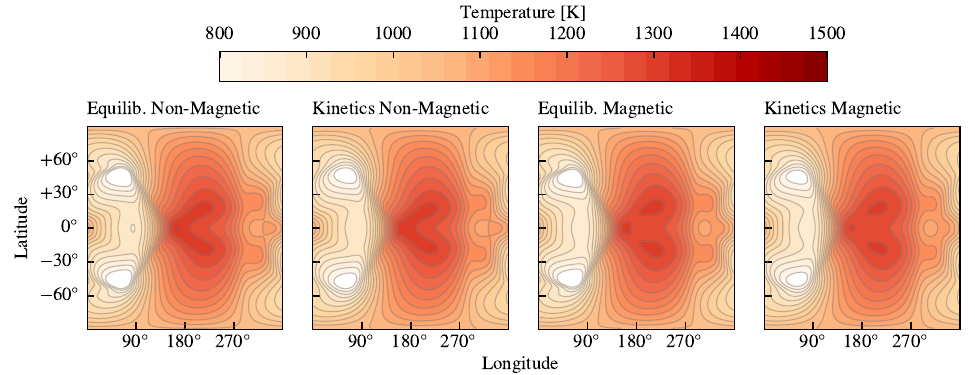}
    \caption{The temperature at 1 mbar for each of the simulations with the fiducial instellation.  In each panel, the substellar point is located in the centre with the anti-stellar point along the left and right edges.} 
    \label{Fig:temp_1x}
\end{figure*}

\begin{figure*}
	\includegraphics[alt={Four panels each with temperature contours for each simulation.}]{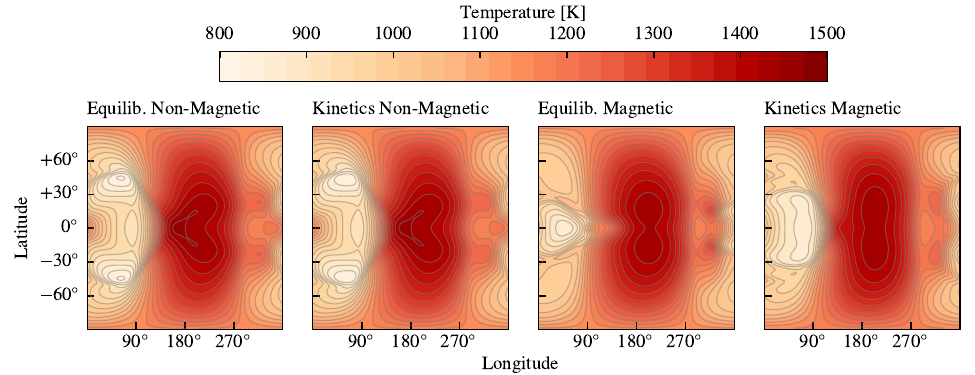}
    \caption{The temperature at 1 mbar for each of the simulations with $1.5\times$ the fiducial instellation. With the increased instellation, the dayside becomes hotter and more ionised.  This results in a suppression of the jet the result of which can be seen in the two rightmost panels where the equatorial nightside is cooler and the dayside no longer exhibits the chevron morphology around the substellar point, located in the center of each panel.} 
    \label{Fig:temp_15x}
\end{figure*}

We first examine the thermal structures of each simulation. The equatorial and mid-latitude profiles,  are found in Figures \ref{Fig:pt_1x} and \ref{Fig:pt_15x}. In the fiducial HD\,209458\,b cases, the chemical equilibrium and kinetics cases exhibit little difference, with the introduction of magnetic drag acting as the only source of discrepancy, although the magnitude is on the order of 30 K. While some deviations of atomic K, Na, and Li from their equilibrium abundances do occur, it is primarily at low pressures on the cooler nightside and early morning (see below) which are less impacted by deviations in abundances of optical absorbers. The increased instellation case similarly shows minimal difference ($\leq 2\,\mathrm{K}$) between the non-magnetic equilibrium and chemical kinetics cases; however, the magnetic drag models exhibit departures from both the non-magnetic models, up to 225 K at 3 bar at the substellar point, as well as from each other, up to 170 K at 0.7 mbar along the morning terminator, as the differing distribution of charged species between the kinetics and equilibrium simulations begins to impact the flow and thus the temperature structure. The largest differences between the magnetic and non-magnetic runs occur deeper in the atmosphere in the increased instellation case due to the jet being slowed by the magnetic drag, altering the rate at which it deposits heat into the deep atmosphere. While the two magnetic runs also show some deep atmosphere differences in the increased instellation case, the largest differences occur at lower pressure where disequilibrium ion abundances alter transport of heat.

The altered temperature structure can also be seen along isobaric surfaces, as in Figures \ref{Fig:temp_1x} and \ref{Fig:temp_15x} which show the temperatures at 1 mbar. In the fiducial ionisation cases, some splitting of the hot spot is seen in the magnetic cases, with the hotspots not being at the equator. In the increased instellation cases, the departure becomes more apparent, both on the dayside and on the nightside, as the increased ionisation on the dayside disrupts the jet which propagates around the planet.

\subsection{Quenching of Chemical Species}
\label{Sec:Quenching}
\begin{figure*}
	\includegraphics[alt={Four panels each with line plots showing the abundances of all potassium bearing species.}]{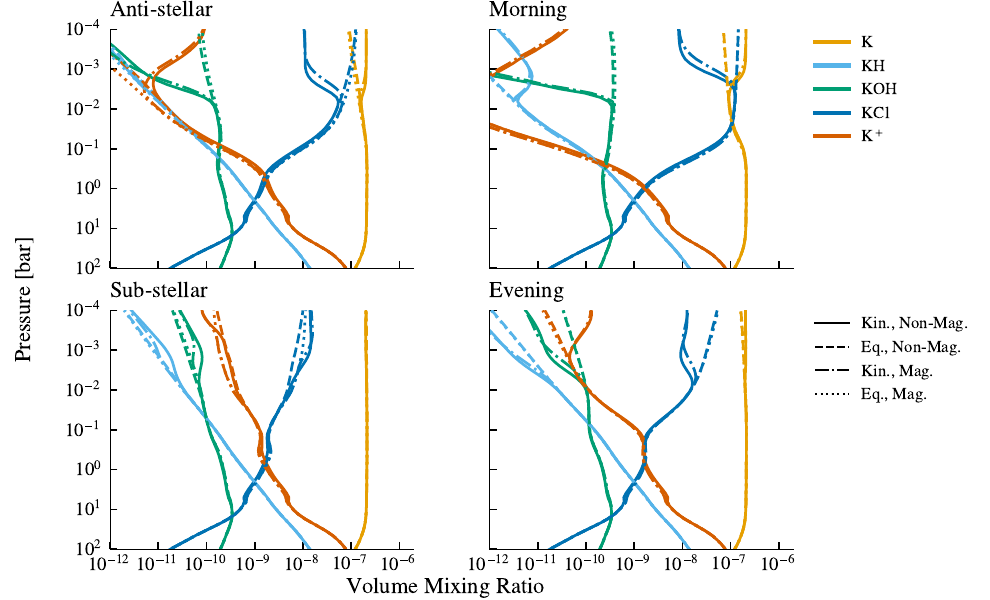}
    \caption{The equatorial abundances of K-bearing species for the fiducial HD~209458~b cases. \ce{K} is mostly atomic, except in the coolest parts of the atmosphere such as the morning terminator.  At pressures less than $\sim 1$ mbar, \ce{K} remains in the atomic form in the chemical kinetics simulations as the chemical timescales exceed the advection timescale within the jet.  The plot for the $1.5\times$ the fiducial instellation case, as well as similar plots for \ce{Na}-bearing species and \ce{Li}-bearing species can be found in Appendix \ref{Appendix:AddPlots}.} 
    \label{Fig:chem_K}
\end{figure*}

For the simulations presented here, the dominant K-bearing species is atomic K except for parts of the nightside and morning terminator at pressures less than $\sim 0.1$ bar. In these regions, the most abundant K-bearing species is \ce{KCl} (see Figure \ref{Fig:chem_K}). At pressures less than $\sim 1$ mbar, horizontal quenching begins to occur, with the interconversion timescale of atomic \ce{K} on the dayside being longer than the advective timescale of the jet, reducing the formation of \ce{KCl} at these pressures. \ce{KOH} and \ce{KH} similarly demonstrate departures from equilibrium abundances at pressures of $1$ to $10$ mbar, although they have neither large abundances nor are they included as opacities, and thus their impact is minimal. The increased instellation cases also demonstrate similar behaviour, although in smaller regions as the higher temperatures result in a larger volume of the atmosphere being primarily atomic K. Na and Li behave similarly, although we do not include figures for the sake of brevity. Quenching of neutral alkali species may have a more dramatic impact for warm Jupiters and Neptunes where in chemical equilibrium a larger volume of the atmosphere is expected to be primarily in the form of alkali-bearing molecules, and transport-induced quenching could both increase the abundance of atomic \ce{K} and \ce{Na} and reduce the abundance of condensible species such as \ce{KCl}. These cooler planets are also more likely to have elemental compositions significantly different than solar, unlike the idealised hot Jupiters investigated here, which could further alter the quenching behaviour. These issues are, however, beyond the scope of this paper.

\begin{figure*}
	\includegraphics[alt={Four panels each with line plots showing the abundances of all charged species.}]{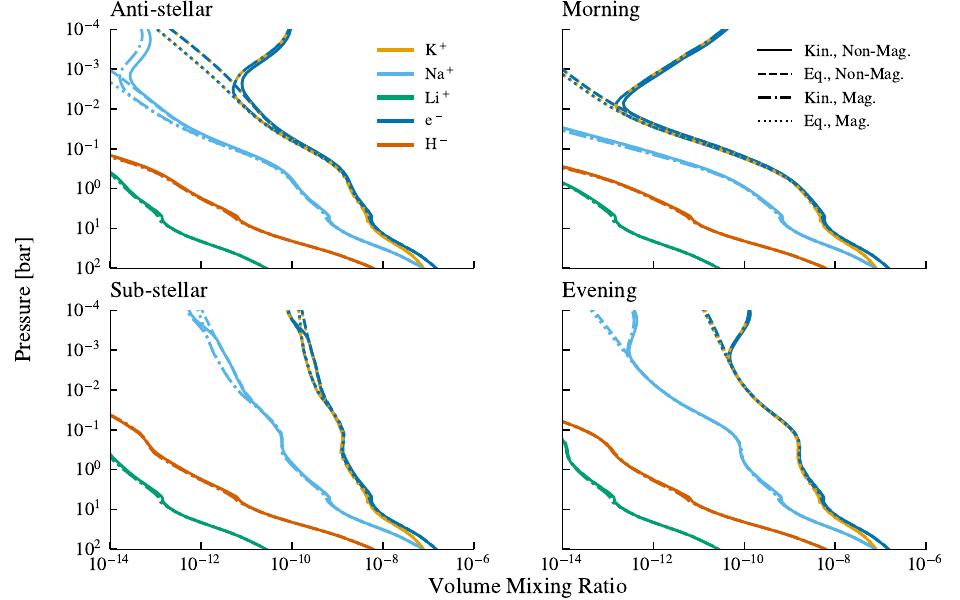}
    \caption{The equatorial ion abundances for the simulations with the fiducial HD~209458~b instellation. Note that the lines for \ce{K+} and \ce{e-} often overlap as \ce{K} atoms are the primary source for electrons. Partial quenching of the charged species begins between 1 and 10 mbar, and it is especially prevalent on the morning terminator. }
    \label{Fig:chem_ion_1x}
\end{figure*}

\begin{figure*}
	\includegraphics[alt={Four panels each with line plots showing the abundances of all charged species.}]{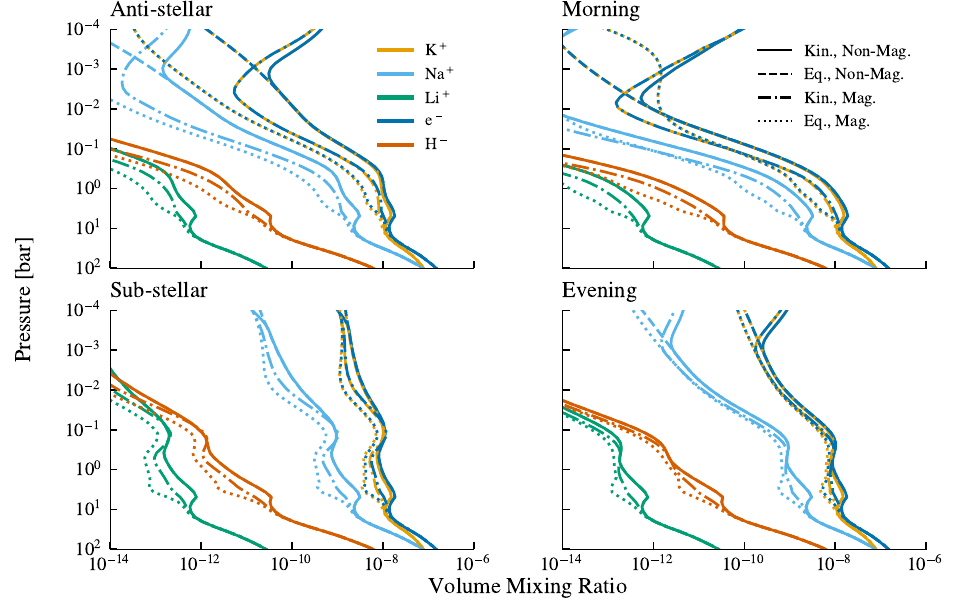}
    \caption{The equatorial ion abundances for the simulations with $1.5\times$ the HD~209458~b instellation. Note that the lines for \ce{K+} and \ce{e-} often overlap as \ce{K} atoms are the primary source for electrons. Due to the magnetic drag altering the circulation, the magnetic chemical kinetics case shows a more complex departure from the equilibrium distribution compared to what is seen in the fiducial instellation cases.}
    \label{Fig:chem_ion_15x}
\end{figure*}

The primary impact of disequilibrium chemistry in the alkali species for the parameters examined here is through their alteration of the ionisation state, relevant for the determination of the drag timescale in the magnetic models. The equatorial abundances of all charged species are shown in Figures \ref{Fig:chem_ion_1x} and \ref{Fig:chem_ion_15x}. For the fiducial instellation cases, departures from equilibrium values occur between 1 and 10 mbar, with the longitudinal dependency characteristic of horizontal quenching: the abundances are close to their equilibrium values at the substellar point where the high temperatures increase the ionisation fraction. The ions are transported across the terminator to the nightside. Three-body recombination remains a fast enough process below 0.1 mbar to avoid the complete homogenisation of the ion abundances, with the profiles instead exhibiting a power-law dependence due to the increasing recombination timescale with decreasing density. This departure from equilibrium is especially apparent for the morning terminator where the atmosphere is the coolest and thus would have the lowest equilibrium thermal ionisation (see Figures \ref{Fig:chem_ion_1x} and \ref{Fig:chem_ion_15x}, upper right panels), resulting in increases in electron abundance by up to five orders of magnitude at 0.1\,mbar depending on the specific simulation parameters. We discuss the potential impact on the resistivity in Section \ref{Sec:Resistivity}.

\begin{figure*}
	\includegraphics[alt={Four panels each with a contour plot showing the distribution of electrons at a fixed pressure.}]{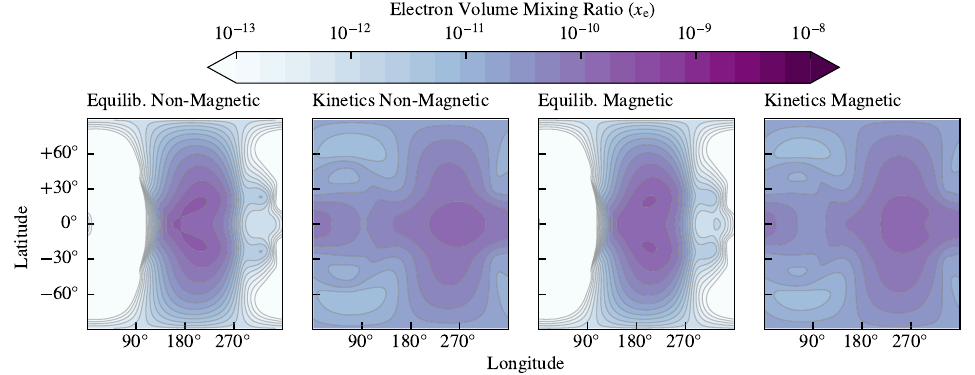}
    \caption{The electron volume mixing ratios $x_\mathrm{e}$ at $0.1$ mbar for each of the simulations with the fiducial HD~209458~b instellation. As \ce{K} is the primary source of electrons and as the gas is charge-neutral, this also traces the \ce{K+} volume mixing ratio. In the chemical equilibrium cases, the cool temperatures result in extremely low $x_\mathrm{e}$ ($\ll 10^{-13}$). In the chemical kinetics cases, the long recombination timescale at these pressures allows for dayside electrons to be transported to the nightside. }
    \label{Fig:horiz_xe_1x}
\end{figure*}

\begin{figure*}
	\includegraphics[alt={Four panels each with a contour plot showing the distribution of electrons at a fixed pressure.}]{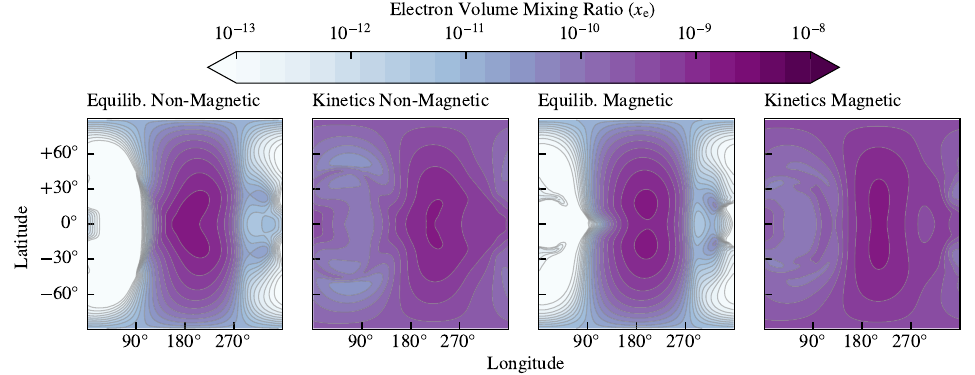}
    \caption{The electron volume mixing ratios $x_\mathrm{e}$ at $0.1$ mbar for each of the simulations with $1.5\times$ the HD~209458~b instellation. As \ce{K} is the primary source of electrons and as the gas is charge-neutral, this also traces the \ce{K+} volume mixing ratio. As in Figure \ref{Fig:horiz_xe_1x}, the long recombination timescale allows for electrons to be transported from the dayside to the nightside in the chemical kinetics cases, only here in larger numbers due to the higher temperatures.}
    \label{Fig:horiz_xe_15x}
\end{figure*}

While Figures \ref{Fig:chem_ion_1x} and \ref{Fig:chem_ion_15x} focus on the equatorial abundances as it best captures the influence of the jet, this departure from equilibrium ionisation abundances occurs at all latitudes, as exhibited in Figures \ref{Fig:horiz_xe_1x} and \ref{Fig:horiz_xe_15x}. This increase of electrons -- and due to charge neutrality, ions -- towards the poles becomes especially relevant as it will increase the magnetic drag not only in the jet but also near the poles and within the nightside gyres, altering the flows there as well.

\begin{figure*}
	\includegraphics[]{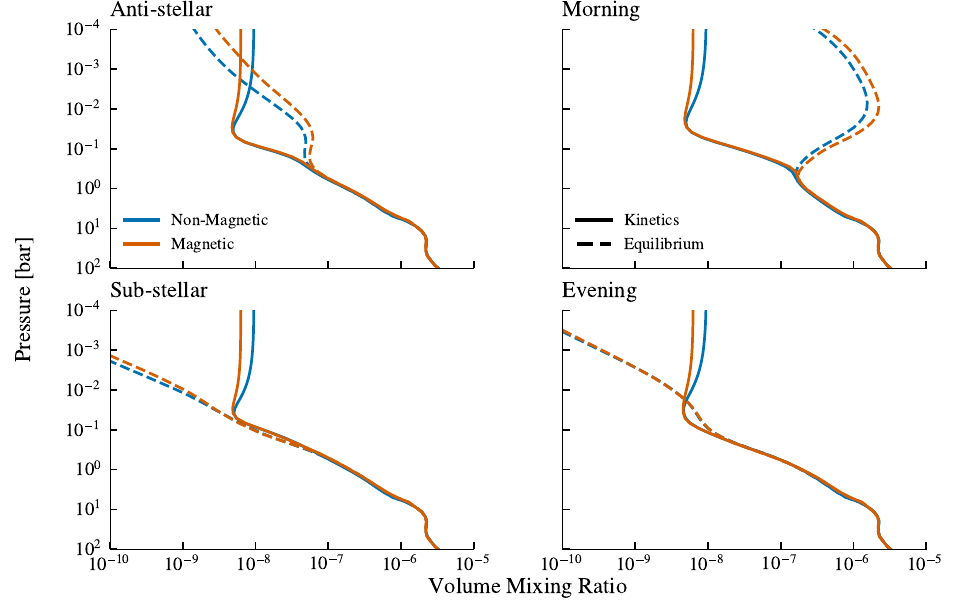}
    \caption{Equatorial \ce{CH4} abundances for each of the simulations with the fiducial HD~209458~b instellation. In the chemical kinetics simulations, \ce{CH4} has a quench point between 10 and 100 mbar with small differences between the magnetic and non-magnetic quenched abundances due to differing equatorial windspeeds.}
    \label{Fig:chem_CH4_1x}
\end{figure*}

\begin{figure*}
	\includegraphics[]{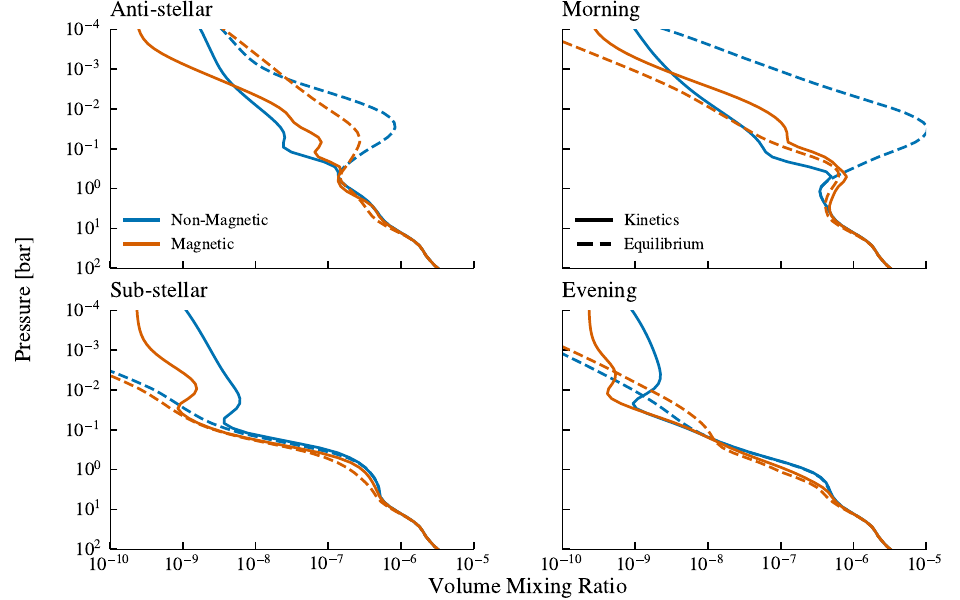}
    \caption{Equatorial \ce{CH4} abundances for each of the simulations with $1.5\times$ the fiducial HD~209458~b instellation. Unlike in the fiducial instellation cases where quenching homogenises \ce{CH4} in the chemical kinetics cases, differences exist between \ce{CH4} profiles on different cardinal directions.  As the atmosphere is hotter and thus more ionised, the impact of magnetic drag on the circulation are larger, and thus larger differences between magnetic and non-magnetic simulations can be observed as well.}
    \label{Fig:chem_CH4_15x}
\end{figure*}

As the magnetic drag begins to alter the circulation, the quenching of non-alkali species can begin to be influenced. This is especially prominent in the case of \ce{CH4}. For the fiducial instellation, the equilibrium simulations show small departures due the temperature differences between magnetic and non-magnetic cases, with similarly small differences in quenching due to both the temperature differences and the altered flow, but the quenched abundances remain relatively homogeneous in longitude (see Figure \ref{Fig:chem_CH4_1x}). For the increased instellation cases, not only are these differences exacerbated due to the larger temperatures and altered flow structures resulting from the magnetic drag, with differences of up to or in excess of an order of magnitude (see Figure \ref{Fig:chem_CH4_15x}), but the quenched \ce{CH4} abundances are no longer homogeneous in longitude, as is evident in the anti-stellar and morning profiles. Furthermore, the quenched abundances in the non-magnetic case are not consistently larger than in the magnetic case (or visa versa), but instead show a more complex behaviour, with the profiles crossing at $\sim$ 1 mbar. This is due to the differing electron and ion distributions altering the flow, as will be discussed in the next section.

\subsection{Impact on Dynamics}

\label{Sec:Dynamics}

\begin{figure*}
	\includegraphics[]{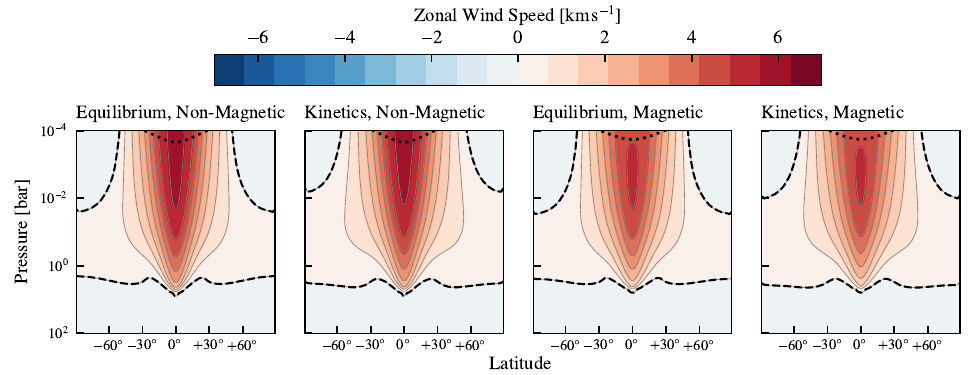}
    \caption{The zonally-averaged wind speed for each of the simulations with the fiducial HD~209458~b instellation. The black dashed line indicates the boundary between super-rotating and counter-rotating regions. The dotted black line indicates where the isobaric surfaces begin to intersect the sponge layer. The inclusion of magnetic drag serves to slow the jet, but is insufficient to dramatically alter the morphology of the flow, thus the qualitative similarities between all four cases.}
    \label{Fig:zonalwind_1}
\end{figure*}

\begin{figure*}
	\includegraphics[]{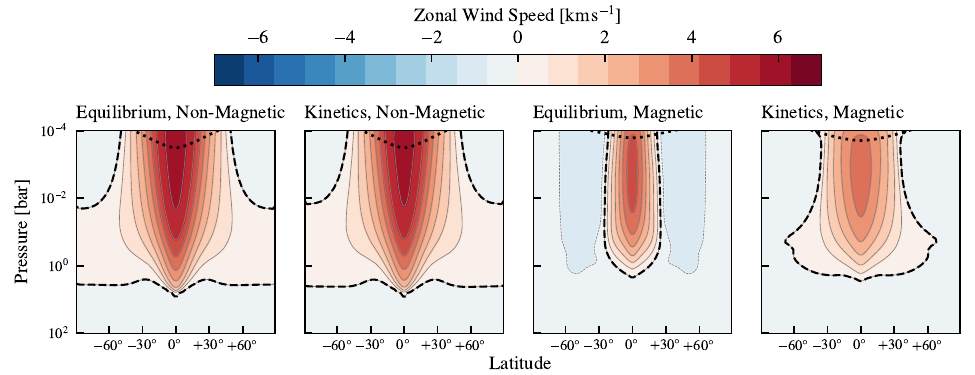}
    \caption{The zonally-averaged wind speed for each of the simulations with $1.5\times$ instellation. The black dashed line indicates the boundary between super-rotating and counter-rotating regions. The dotted black line indicates where the isobaric surfaces begin to intersect the sponge layer. The simulations including magnetic drag show larger differences compared to the fiducial instellation cases. In the equilibrium magnetic drag case, the drag remains localised near the equator, allowing for a faster counter-rotating flow at the mid-latitudes in the nightside gyres.  In the chemical kinetics case, the electrons and thus the drag are more uniformly distributed, slowing the circulation globally.}
    \label{Fig:zonalwind_1_5}
\end{figure*}

\begin{figure}
	\includegraphics[]{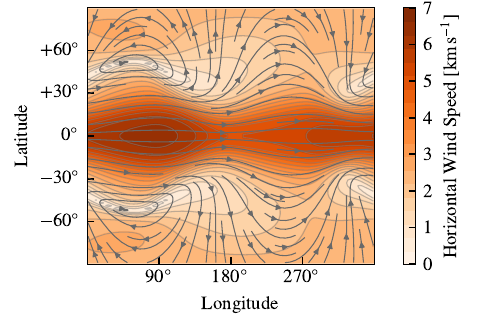}
    \caption{The horizontal wind speed at 1 mbar for the fiducial instellation case in chemical equilibrium without magnetic drag included. As the velocity field is largely the same for all simulations with this instellation, we omit the remaining plots as they are qualitatively very similar, with the magnetic models exhibiting slightly slower wind speeds.}
    \label{Fig:horiz_vel_1x}
\end{figure}

\begin{figure*}
	\includegraphics[]{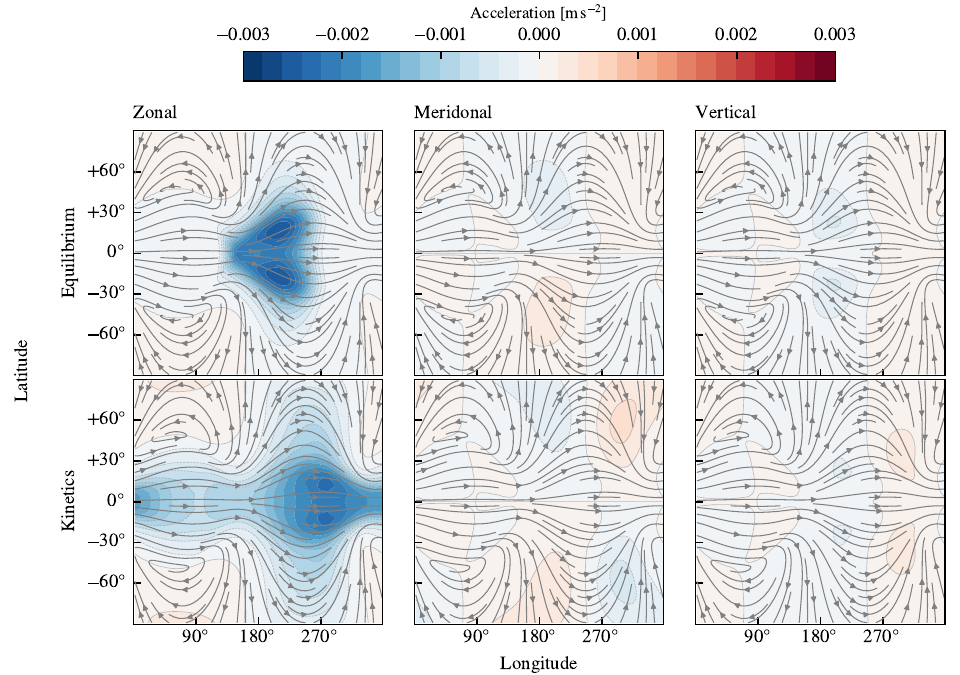}
    \caption{The zonal (left), meridional (middle), and vertical (right) components of the local magnetic drag for the fiducial HD209458b case at 0.1 mbar. The equilibrium simulation is shown in the top row and the chemical kinetics simulation in the bottom row. The drag accelerations are shown at a pressure of 0.1 mbar to highlight the departure of the kinetics case from equilibrium case. As the circulation is dominated by the jet, the largest drag remains zonal drag near the equator, with insufficient electron abundance at mid-latitudes the impact the flow there. }
    \label{Fig:drag_1x_01mbar}
\end{figure*}

\begin{figure*}
	\includegraphics[]{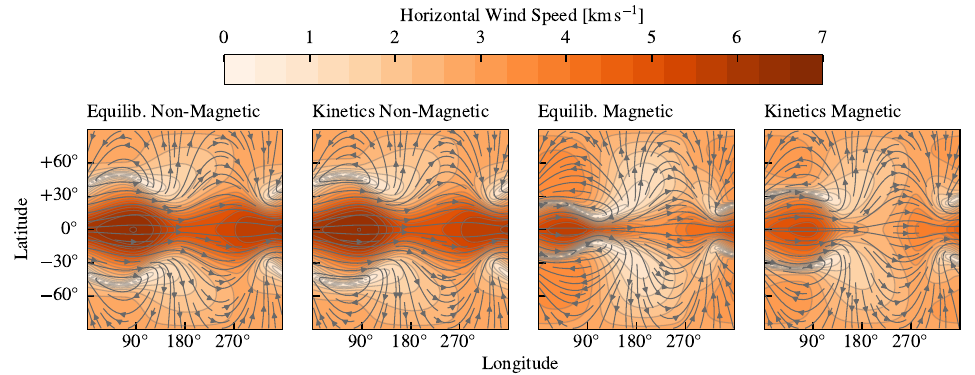}
    \caption{The horizontal wind speed at 1 mbar for all simulations with $1.5\times$ the fiducial instellation. While the two non-magnetic drag simulations have similar jet morphologies due to the minimal impact of chemical kinetics in the non-magnetic simulations, the magnetic drag cases differ in width of the jet and the equatorial wind speeds across the dayside.  The fast counter-rotating zonal flow at mid-latitudes seen in the equilibrium magnetic drag case in Figure \ref{Fig:zonalwind_1_5} can also be seen here in the upper left of the sescond panel from the right.}
    \label{Fig:horiz_vel_1.5x}
\end{figure*}

The mean zonal velocities are shown in Figures \ref{Fig:zonalwind_1} and \ref{Fig:zonalwind_1_5}. For the fiducial case, we find that both non-magnetic simulations are almost identical with peak mean zonal wind speeds of 5.7 to 5.8 km s$^{-1}$, with the magnetic simulations exhibiting a slower jet ($\sim$ 5 km s$^{-1}$). Although the winds are slightly slower in the magnetic simulations, the simulations with the fiducial instellation are all morphologically similar (see Figure \ref{Fig:horiz_vel_1x}). While the magnetic simulations have similar zonal velocities, the zonal forcing at low pressures ($\sim$ 0.1 mbar; see Figure \ref{Fig:drag_1x_01mbar}) extends throughout the jet in the chemical kinetics cases whereas it is localized near the substellar point in the equilibrium case. Neither case shows significant meridional or vertical drag. Since the distribution of the forcing is different between the magnetic cases, it is possible that the zonal velocities may diverge over integration times longer than the 1000 Earth days adopted here. At the increased instellation, we find similar mean zonal wind speeds for the non-magnetic cases ($\sim$ 5.8 km s$^{-1}$); however, for the magnetic cases, the peak mean zonal winds slow to 4.5 km s$^{-1}$ in the equilibrium chemistry case and 3.9 km s$^{-1}$ in the chemical kinetics case. In addition to the slowing of the equatorial jet, the magnetic models exhibit different behaviours in the counter-rotating flow at mid-latitudes (see the right two panels of Figure \ref{Fig:zonalwind_1_5} as well as Figure \ref{Fig:horiz_vel_1.5x}). In the equilibrium case, electrons and ions track the temperature and are concentrated on the dayside, especially near the equator while in the chemical kinetics case the increased ionisation on the nightside and at mid-laitudes increases the zonal drag within the counter-rotating nightside gyres but also allows for increased drag at mid-latitudes on the dayside. We further note that in the increased instellation case the increased ionisation in the poleward flows resulting from chemical kinetics allows for increased meridional drag, in addition to the zonal drag associated with the super-rotating equatorial jet. These differences in flow demonstrate the importance of ion quenching in shaping the atmospheric circulation.

\begin{figure*}
	\includegraphics[]{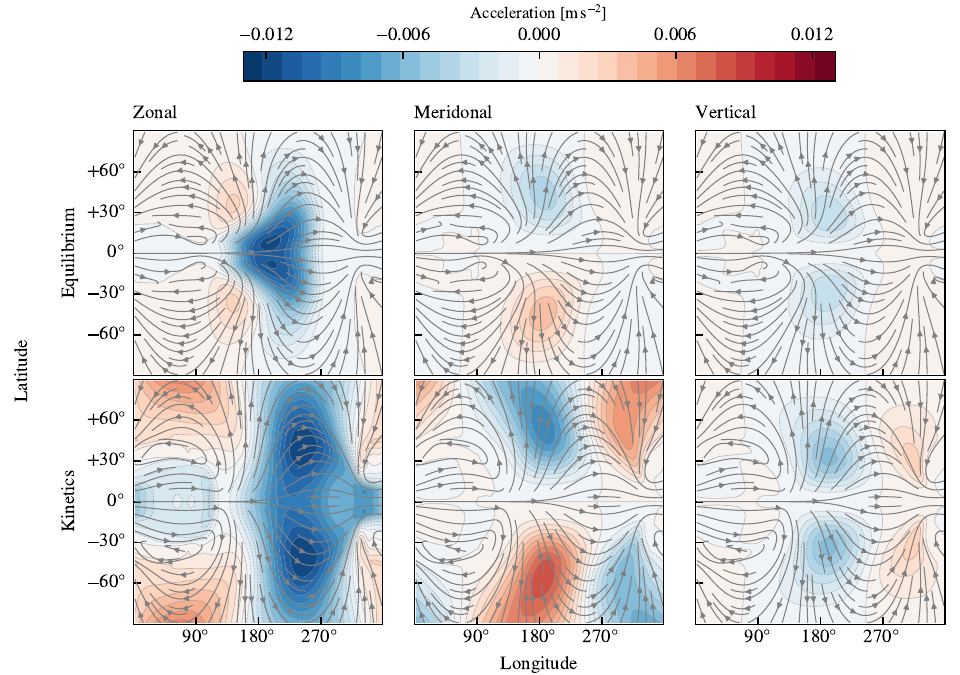}
    \caption{The zonal (left), meridional (middle), and vertical (right) components of the local magnetic drag for the $1.5\times$ instellation case at 0.1 mbar. The equilibrium simulation is shown in the top row and the chemical kinetics simulation in the bottom row. The drag accelerations are shown at a pressure of 0.1 mbar to highlight the departure of the kinetics case from equilibrium case. Note that the colour scale is different from Figure \ref{Fig:drag_1x_01mbar}. The greater impact of chemical kinetics can be observed here, as electrons are able to be transported to the mid-latitudes and towards the poles, increasing both the meridional drag as well as the zonal drag within the counter-rotating gyres.}
    \label{Fig:drag_1_5x_01mbar}
\end{figure*}

\subsubsection{Atmospheric Torque}

 Within the magnetic drag formalism, there is no constraint on the torque on the atmosphere within the computational domain to be zero, and \citet{wazny_2025} have argued that the atmospheric torques due to magnetic drag must exert a commensurate torque on the interior, assuming no angular momentum is lost through the upper boundary. Within the context of their drag model, \citet{christie_2025} found a net non-zero torque as well, although the magnitude was smaller than found by \citet{wazny_2025} due to the self-consistent application of magnetic drag and the spatial dependence of the magnetic field strength in the \citet{christie_2025} simulations. As chemical kinetics alters the distribution of charged species, the magnetic drag forces are similarly altered, allowing for the possibility of further change of the net torque. We compute the torque due to magnetic drag $\Gamma_\mathrm{drag}$ as

 \begin{equation}
\Gamma_\mathrm{drag} = \int_V dV \rho \left(\frac{\partial u}{\partial t}\right)_\mathrm{magnetic} r\cos\theta = -\int_V dV \frac{\rho u}{\tau_\mathrm{drag}}r\cos\theta \,\, ,
\label{Eqn:torque}
\end{equation}

\noindent where $\rho$ is the gas density, $r$ is the radial coordinate with the origin at the centre of the planet, and the integral is over the computational domain $V$. The sign convention adopted here is that a negative net torque is opposing the super-rotating jet while a positive torque reinforces it. The resulting torques for each simulation are summarized in Table \ref{Tbl:Torque}. The magnitude of the torques and the signs are due to a confluence of modelling choices. In their initial analysis, \citet{wazny_2025} found, in post-processing their non-magnetic simulations with a torque estimate based on magnetic drag assuming a constant magnetic field strength, that the modelled atmosphere experienced a torque opposing the equatorial jet.  In an examination of their self-consistent magnetic drag simulations using a local varying magnetic field strength, \citet{christie_2025} found that while the drag on the equatorial jet did exert a negative torque on the atmosphere, the reduced field strength and thus reduced drag in the upper atmosphere resulted in the magnitude of this torque being less than the opposing torque on counter rotating jets deeper in the atmosphere.   This results in a net positive torque on the atmosphere (see Figure 18 of \citealt{christie_2025}).  We find that for the equilibrium chemistry simulations performed here the results are consistent with the previous results of \citet{christie_2025}, with the torque on the counter-rotating jets deep the atmosphere outweighing the torque on the equatorial jet higher in the atmosphere.  This interpretation also holds for the chemical kinetics simulations in the fiducial instellation case, as magnetic forcing in the upper atmosphere occurs primarily within the jet (see Figure \ref{Fig:drag_1x_01mbar}). The chemical kinetics case with $1.5\times$ the fiducial instellation, on the other hand, has increased forcing in the upper atmosphere across the dayside compared to the equilibrium case (see Figure \ref{Fig:drag_1_5x_01mbar}), resulting in a larger fraction of the upper atmosphere experiencing a negative torque.  The result is that in this case negative torque in the upper atmosphere outweighs the positive torque on the deep atmosphere, opposite to what is found in the equilibrium case. We conclude that, as in \citet{christie_2025},  modelling choices greatly alter the atmospheric torque, making the quantification impact of angular momentum redistribution to the interior or exterior difficult.  


\begin{table}
\caption{Atmospheric Torque ($\times 10^{23}\,\,\mathrm{N\,m}$)}
\label{Tbl:Torque}
\begin{tabular}{lcc}
\hline
\hline
  & Fiducial & $1.5\times$ Instellation\\
\hline
Equilibrium        & $2.699$ & $30.75$ \\
Chemical Kinetics  & $1.996$ & $-21.03$ \\
\hline
\end{tabular}
\end{table}

\subsection{Observational Impact}

To understand the extent that the inclusion of alkali chemical kinetics and magnetic drag can alter observables, we generate phase curves and transmission spectra for each of the simulations using the UM's {\sc socrates} radiative transfer module.  While these synthetic observables are created at a higher spectral resolution, they use the same opacity sources and radiative transfer scheme as the simulations themselves. 

\subsubsection{Transmission Spectra}

\begin{figure*}
	\includegraphics[]{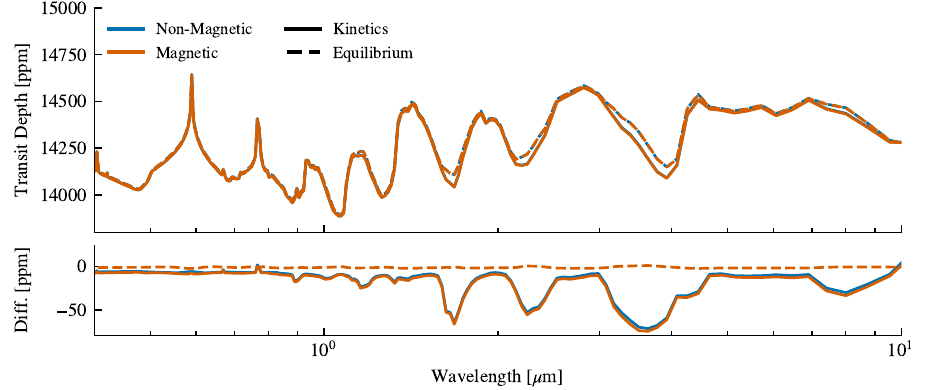}
    \caption{A comparison of the full limb transmission spectra for the fiducial instellation cases. {\em Top:} The synthetic transmission spectra for each of the simulations. {\em Bottom:} The difference between a simulation's transmission spectrum and that of the non-magnetic, equilibrium case. The differences between the equilibrium and chemical kinetics cases are due to the quenching and homogenisation of \ce{CH4} as discussed in Section \ref{Sec:Quenching} and shown in Figure \ref{Fig:chem_CH4_1x}. } 
    \label{Fig:trans_1x}
\end{figure*}

\begin{figure*}
	\includegraphics[]{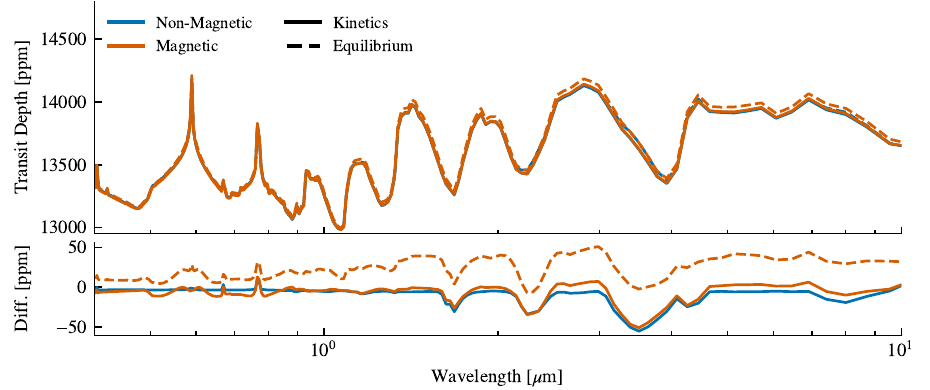}
    \caption{A comparison of the full limb transmission spectra for the $1.5\times$ fiducial instellation cases. {\em Top:} The synthetic transmission spectra for each of the simulations. {\em Bottom:} The difference between a simulation's transmission spectrum and that of the non-magnetic, equilibrium case. While differences due to \ce{CH4} abundances are observed as in the fiducial instellation cases (see Figure \ref{Fig:trans_1x}), additional differences are due to differences in the thermal structure due to magnetic drag altering the circulation. } 
    \label{Fig:trans_15x}
\end{figure*}

The details of the implementation of the UM's internal transmission spectra diagnostics are outlined in \citet{lines_2018b}. The transmission spectra are computed mid-transit on a non-uniform, 500-wavelength grid from 0.2 $\mu$m to 100 $\mu$m with $R=\lambda/\Delta \lambda \sim 100$ at $\lambda=1\,\mu$m. The synthetic spectra for each simulation are shown in Figures \ref{Fig:trans_1x} and \ref{Fig:trans_15x} as well the relative difference between a spectra and the spectrum from the non-magnetic equilibrium simulation with the same instellation. In the fiducial instellation case, the differences of up to 75 ppm arise from the quenching behaviour of \ce{CH4} with the magnetic and non-magnetic equilibrium simulations only differing by at most 3 ppm. In the increased instellation case, we see some differences ($\sim$ 55 ppm relative to the equilibrium non-magnetic simulation) in the spectra due to the quenching behaviour of \ce{CH4} in the chemical kinetics simulations (see Section \ref{Sec:Quenching}); however, the largest factor is the differing temperature structure in the equilibrium magnetic simulation, resulting in a difference of up to 51 ppm relative to the non-magnetic equilibrium simulation.

\begin{figure*}
	\includegraphics[]{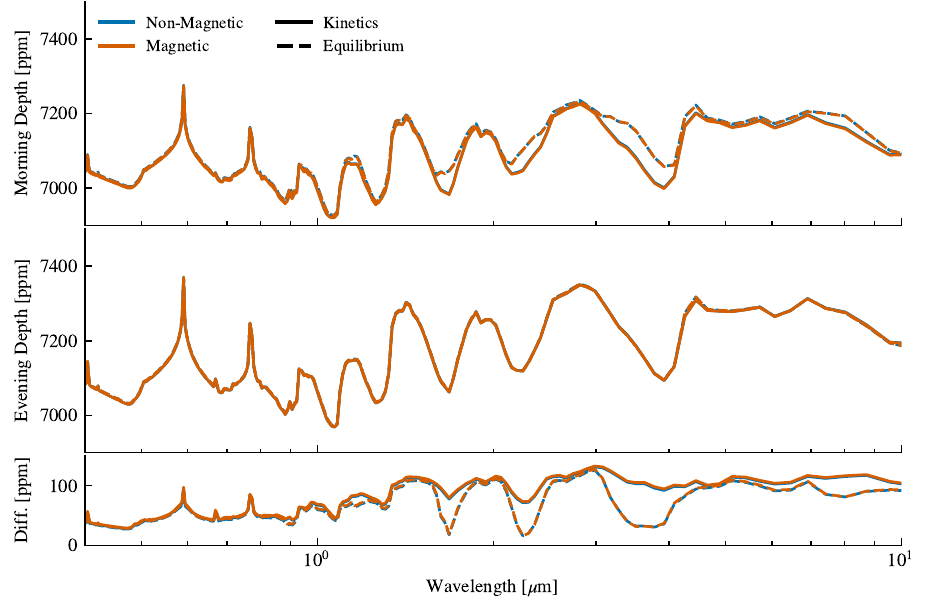}
    \caption{A comparison of the morning and evening limb transmission spectra for the fiducial instellation cases. {\em Top:} The synthetic transmission spectra for the morning limb for each of the simulations. {\em Middle: } The synthetic transmission spectra for the evening limb for each of the simulations. {\em Bottom:} The difference between a simulation's evening and morning limbs, with a positive value representing a larger evening limb. } 
    \label{Fig:trans_limb_1x}
\end{figure*}

\begin{figure*}
	\includegraphics[]{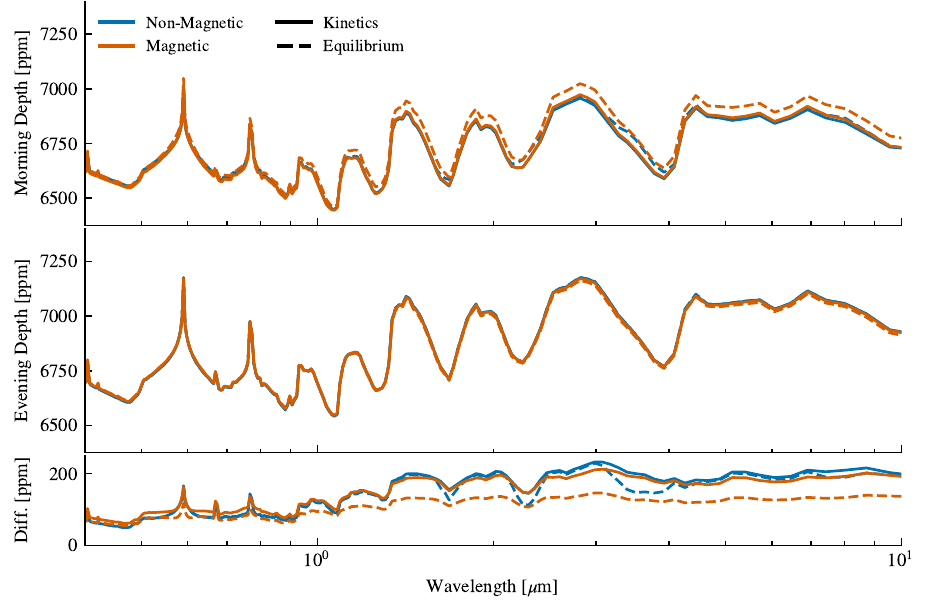}
    \caption{A comparison of the morning and evening limb transmission spectra for the $1.5\times$ fiducial instellation cases. {\em Top:} The synthetic transmission spectra for each of the simulations. {\em Middle: } The synthetic transmission spectra for the evening limb for each of the simulations. {\em Bottom:} The difference between a simulation's evening and morning limbs, with a positive value representing a larger evening limb.  } 
    \label{Fig:trans_limb_15x}
\end{figure*}

The individual limbs (Figures \ref{Fig:trans_limb_1x} and \ref{Fig:trans_limb_15x}) exhibit asymmetry,  with every simulation having a larger evening limb relative to the morning limb.   In the fiducial instellation case (Figures \ref{Fig:trans_limb_1x}), the evening limbs show minimal variation between simulations while the spectra of the morning limbs show differences between the chemical equilibrium and chemical kinetics cases.  The limb asymmetries are up to 133 ppm, with the asymmetries between the evening and morning spectra primarily being due to the temperature differences between the limbs.  In the chemical equilibrium cases, however, the morning limbs have increased \ce{CH4} abundance (Figure \ref{Fig:chem_CH4_1x}) and this additional absorption serves to reduce the asymmetry between the limbs in regions around \ce{CH4} absorption features.  As chemical kinetics zonally homogenises the \ce{CH4} abundance, this is not observed in the limb spectra for these simulations.  

The increased instellation cases exhibit larger limb asymmetries (up to 233 ppm), as well as larger spectral differences on the morning limb between different simulations. This is especially prominent in the chemical equilibrium magnetic drag case, as in this simulation magnetic drag had the largest effect in disrupting the jet (see Figure \ref{Fig:horiz_vel_1.5x}), and thus altering the thermal structure on the morning limb.

The increased instellation cases also exhibit differences in absorption of $\sim$ 20 ppm in the \ce{K} line at 0.766 $\mu$m (Figure  \ref{Fig:trans_15x}). While a relatively minor difference due to quenching, we note that potassium is clearly visible in both simulations, and a lack of detectable neutral potassium in a hot Jupiter spectrum may point towards photoionisation depleting the neutral atomic population or a reduced elemental abundance of potassium. These two potential reasons for an unobserved potassium line in an aerosol-free atmosphere would have opposing impacts on the dynamics. Photoionisation depleting the neutral atomic population \citep{fortney_2003} would greatly increase the electron abundance, at least on the dayside, resulting in shorter magnetic drag timescales and a field likely prone to deformation by the winds. An elemental depletion of potassium would remove the primary source of electrons in hot Jupiter atmospheres, reducing the electron abundance and further decoupling the dayside magnetic field from the atmosphere, limiting the importance of magnetic effects in regulating the global circulation. Determining which reflects reality requires a proper accounting of the elemental potassium in the atmosphere, and if it is not directly observable, an understanding how how it may be impacted by formation and evolutionary processes.

\subsubsection{Phase Curve}

\begin{figure*}
	\includegraphics[]{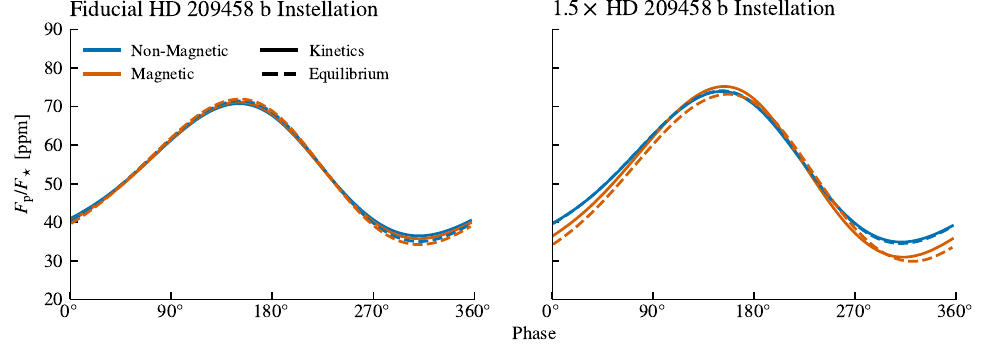}
    \caption{The synthetic phase curves for each of the simulations, including all flux between 0.2 $\mathrm{\mu}$m and 200 $\mathrm{\mu}$m. The inclusion of magnetic drag reduces the heat redistribution between the dayside and nightside, decreasing the nightside flux, especially in the $1.5\times$ HD~209458~b instellation case.   } 
    \label{Fig:phasecurve}
\end{figure*}

We generate phase curves for each of the simulations using the {\sc socrates} diagnostic outputs (see Figure \ref{Fig:phasecurve}), including all flux between 0.2 microns and 200 microns.  In the fiducial instellation case, the differences are on the order of a few ppm and negligible change in the phase offset between simulations, with an offset of $30.5^\circ$ in both cases, consistent with the minor changes in the thermal structure (see Section \ref{Sec:Thermal} above). The increased instellation cases exhibit larger differences, primarily on the nightside, due to magnetic drag reducing the redistribution of heat through the suppression of the jet.  The equilibrium and kinetics non-magnetic simulations have only minor differences, with instead the largest differences being between the magnetic and non-magnetic cases, reaching up to 6 ppm.  The non-magnetic simulations again do not exhibit variation in the phase offset ($30.5^\circ$\footnote{The emission spectra used in creating the phase curves are calculated with a one hour cadence, which limits the phase offset resolution to 4.3$^\circ$. While the phase offsets for six of the simulations are, at this resolution, identically, any inter-simulation differences smaller than this may exist.}) while the magnetic equilibrium and kinetic simulations have phase offsets of $21.9^\circ$ and $26.2^\circ$, respectively.  While the non-magnetic cases exhibit both faster jets and larger phase offsets, as expected, the trend of faster jets corresponding to larger phase offsets is not found when comparing between equilibrium and chemical kinetics cases with magnetic drag, as the altered dayside drag between the equilibrium and kinetics simulations complicates the dayside thermal structure (see Section \ref{Sec:Thermal}).

\subsection{Implication for MHD Models}

Magnetic drag models are employed as an alternative to solving the equations of non-ideal MHD directly in regimes where they are intractable.  We discuss here the possibility of decreasing the nightside resistivity through the inclusion of alkali chemical kinetics as well as the applicability of the magnetic drag approximation in both the equilibrium and chemical kinetics cases.

\subsubsection{Average Resistivity Profile}
\label{Sec:Resistivity}

Models attempting to solve the equations of magnetohydrodynamics for hot Jupiters (i.e., those not employing the magnetic drag approximation) often make assumptions about the resistivity or the ionisation state to avoid having regions of extremely high resistivity which can make the problem numerically intractable. Both \citet{rogers_2014} and \citet{rogers_2014b}, for example, adopt a resistivity profile based on a 1D temperature reference profile, and while the equilibrium temperature varies horizontally, the resistivity does not. \citet{batygin_2013} adopt a further simplification of a constant resistivity throughout the computational domain. Only in the modelling of ultra-hot Jupiters in \citet{rogers_2017} and \citet{rogers_2017b} has this assumption been relaxed, allowing for a horizontally varying resistivity, although the nightside temperatures in these models are $\sim 1800\,\mathrm{K}$, hotter than the dayside temperatures in models examined here.

\begin{figure*}
	\includegraphics[]{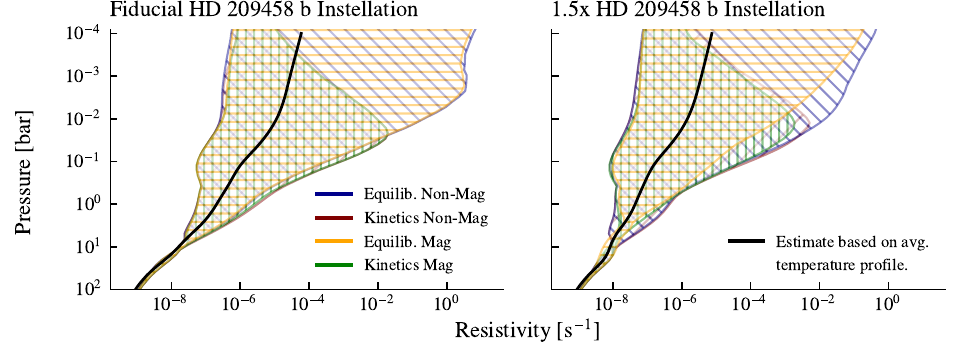}
    \caption{The range of values for the resistivity for all simulations. The solid black line indicates resistivity estimated from the equilibrium electron volume mixing ratio computed using the area-weighted temperature profile. While transport in the chemical kinetics cases leads to an increase in the nightside electron abundances and thus reduces    the resistivity, the resistivity remains large on the nightside, peaking between 10 and 100 mbar.  Furthermore, adopting an resistivity profile based on an average temperature would underestimate the nightside resistivity by up to four orders of magnitude. }
    \label{Fig:etasummary}
\end{figure*}

To understand if horizontal quenching of electrons can justify the assumption of horizontally uniform resistivity, we examine the range of resistivities (Figure \ref{Fig:etasummary}) on isobaric surfaces. We also compute an average temperature profile from the equilibrium non-magnetic simulations and use those profiles to compute an effective resistivity as a function of pressure, assuming equilibrium abundances of electrons and ions. In our equilibrium chemistry simulations the resistivity varies by roughly six orders of magnitude at low pressures, with the resistivity associated with the average temperature still four to five orders of magnitude smaller than the higher resistivities found on the nightside. In the chemical kinetics models, the electron and ion abundances begin to quench, reducing the resistivity on the nightside. This change begins at between 10 and 100 mbar, where the largest resistivities are found, with the largest resistivites on any isobaric surface decreasing with pressure above the maximum. Between 10 and 100 mbar, the resistivity from the mean temperature profile is three orders of magnitude less than the value computed from the chemical kinetics simulation, indicating that resistivities generated from the mean temperature profile do not adequately capture the nightside decoupling throughout the atmosphere. We do, however, note that at pressures of 0.1 mbar in the kinetics simulations, the electrons and ions become sufficiently abundant throughout the atmosphere that the largest resistivity is actually smaller than the resistivity computed from the average temperature. This reduction in the range of resistivities hints that at these lower pressures the issues of nightside resistivities may be mitigated by quenching.

In their comparable MHD model, \citet{rogers_2014b} adopted a horizontally uniform resistivity from a reference state with the resistivity varying between $\sim 10^{-7}$ and $10^{-5}\,\mathrm{s^{-1}}$ for their coolest case (see their Figure 1, also note that the definition of resistivity used here differs by a factor of $c^2/4\pi$). Thus, the resistivities used here are up to three orders of magnitude larger in the kinetics cases and five orders of magnitude larger in the equilibrium cases relative to those of \citet{rogers_2014b}. While what constitutes a tractable problem is subjective and depends on the ability of the MHD scheme to handle large resistivities, the availability of computational resources, as well as the ability of the user to endure long integration times, this does seem to imply that for a HD~209458~b-like planet, quenching of electrons will not make the modelling of MHD more tractable at pressures characteristic of GCMs.

\subsubsection{The Magnetic Reynolds Number}

\begin{figure*}
	\includegraphics[]{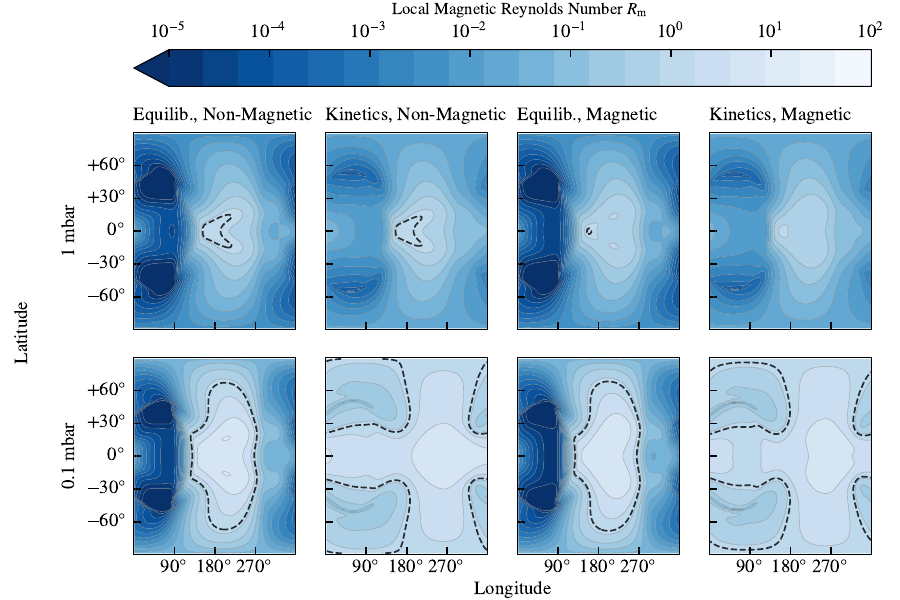}
    \caption{The local magnetic Reynolds number for simulations with the fiducial instellation. The black dashed line indicates the $R_\mathrm{m} = 1$ surface. At 1 mbar, all simulations have $R_\mathrm{m} < 1$ across much of the isobaric surface, indicating that this region is potentially dominated by the diffusive magnetic effects; however, at 0.1 mbar, much of the dayside now satisfies $R_\mathrm{m} > 1$, as does part of the nightside is the chemical kinetics cases, hinting that the magnetic field may begin to deform in these regions (see Section \ref{Sec:LimitMag}). }
    \label{Fig:Rm1x}
\end{figure*}

\begin{figure*}
	\includegraphics[]{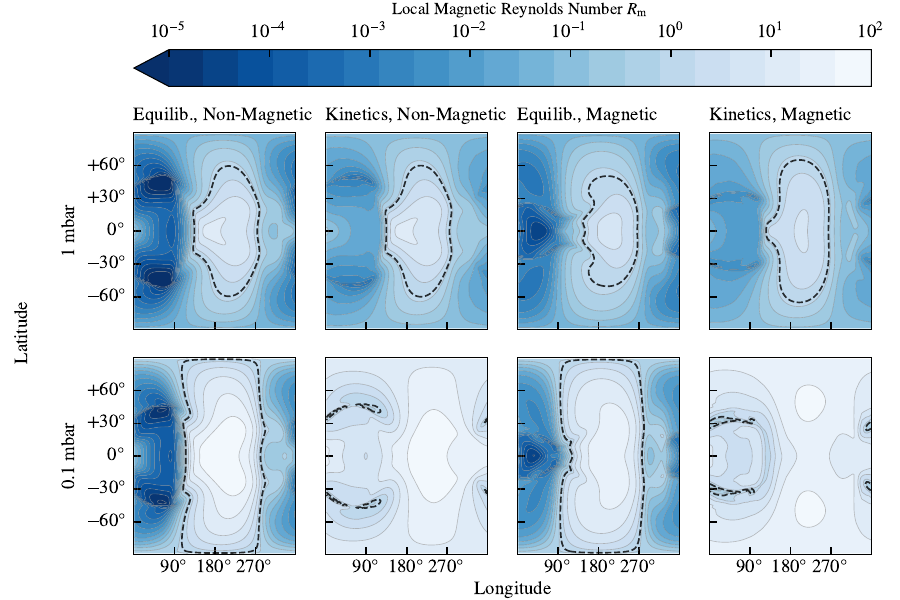}
    \caption{The local magnetic Reynolds number for simulations with $1.5\times$ the fiducial instellation. The black dashed line indicates the $R_\mathrm{m} = 1$ surface. At both pressure levels shown, much of the dayside satisfies $R_\mathrm{m} >1$ while at 0.1 mbar the chemical kinetics simulations have $R_\mathrm{m} > 1$ for almost all of the nightside as well.  As a result, caution should be used in interpreting results as the potential for field deformation exists, as discussed in Section \ref{Sec:LimitMag}. }
    \label{Fig:Rm1.5x}
\end{figure*}

As the nightside resistivity may make integration of the MHD equations difficult, even when the chemical kinetics of the ions and electrons are considered, we return to the applicability of the magnetic drag approximation. Figures \ref{Fig:Rm1x} and \ref{Fig:Rm1.5x} show the magnetic Reynolds number (Equation \ref{Eqn:Rm}) for all simulations, both at 1 mbar, roughly where quenching begins, and at 0.1 mbar, where quenching can be seen to have a more dramatic effect. At 1 mbar in both the non-magnetic and magnetic equilibrium cases, the nightside ionisation is sufficiently low that the magnetic Reynolds number is significantly less than unity, indicating that the nightside magnetic field likely remains decoupled from the atmosphere, where as the dayside has a localized region of $R_\mathrm{m} > 1$ around the substellar point. This region is very small at 1 mbar for the fiducial instellation but it extended in the increased instellation case. At lower pressures, these regions expand, and in the increased instellation case, occupy the entire dayside with the peak values reaching $\sim 50-100$, indicating that some deformation of the field is possible in these regions. As the nightside is most likely decoupled, however, it is unclear to what extent the field can wrap around the planet, as has been discussed in previous works \citep[e.g.,][]{rauscher_2013,beltz_2022,christie_2025}.

In the chemical kinetics cases, the ionisation fraction on the nightside at 1 mbar is increased, which is especially noticeable around the cold, low ionisation centres of the nightside gyres (see Figures \ref{Fig:Rm1x} and \ref{Fig:Rm1.5x}). The magnetic Reynolds numbers are still less than unity on the nightside, with the dayside still having a region of $R_\mathrm{m} > 1$. At the lower pressures of 1 mbar, the magnetic Reynolds number is greater than unity or approaching unity throughout most of the atmosphere. At these pressures, there is again potential for deformation of the field. Combining this with the results deeper in the atmosphere, a picture develops of a nightside atmosphere at pressures higher than 0.1 mbar capable of flowing through the field but in the lower pressure atmosphere above this region the atmosphere begins to deform the field.

We stress the importance that the analysis of the magnetic Reynolds numbers provide the same conclusions when using results from magnetic and non-magnetic simulations, providing a modicum of confidence that the application of magnetic drag is not biasing the conclusions about the relative importance of advection and diffusion by artificially damping the velocity field. That said, we note that in all our simulations here at least part of the dayside satisfied $R_\mathrm{m}>1$, indicating that in the crudest terms that the advection and deformation of the field should be expected, in tension with the underlying assumptions of the magnetic drag models. When chemical kinetics are employed, an even larger fraction of the atmosphere satisfies $R_\mathrm{m}>1$. As the assumptions of the magnetic drag model are already beginning to be bent, if not outright broken, this should give pause before over-interpreting quantitative results using these models, especially for hotter atmospheres.

\section{Discussion and Conclusions}
\label{Sec:Conclusions}

Motivated to better understand magnetic effects in hot Jupiter atmospheres, we have presented simulations of hot Jupiter atmospheres including modelling of the alkali chemistry and demonstrated that alkali ions begin to experience partial horizontal quenching at pressures less than 1-10 mbar. This results in increased electron and ion abundances towards the poles and on the nightside which has the potential to alter the circulation compared to magnetic drag simulations employing equilibrium chemistry. In the increased instellation cases investigated here, the switch from equilibrium chemistry to chemical kinetics resulted in the peak mean zonal wind speed being reduced from 4.5 km\, s$^{-1}$ to 3.9 km\, s$^{-1}$ with increased drag near the poles and in the nightside gyres as electrons and ions are carried away from the hottest points on the dayside. This transport of charged species alters the winds beyond the equator and the net torque on the atmosphere, especially in the increased instellation cases investigated here. Furthermore, as wind speeds derived from spectroscopic observations are being used to infer magnetic field strengths based on drag models \citep{seidel_2026}, capturing the correct electron and ion distribution is essential to accurately interpreting the results. While we model hot Jupiters here and much of the recent focus of magnetic models has instead been on ultra-hot Jupiters \citep[e.g.][]{frazier_2026,fecanin_2026} where the dayside chemical timescales may be sufficient to keep at least part of the dayside in thermochemical equilibrium, the nightside temperatures may still approach $\sim$ 1000 K (see e.g. \citealt{evans-soma_2025} for the case of WASP-121~b and the GCM model of the same planet in \citealt{parmentier_2018}), sufficiently cool to allow transport of ions from the dayside or terminator to the nightside, increasing the nightside ionisation fraction, at least at low pressures. Thus models of ultra-hot Jupiters, even though not investigated here, are likely to require the consideration of the ion chemical kinetics as well to properly understand the nightside dynamics.

We also examined the possibility that the inclusion of chemical kinetics of ions and electrons could sufficiently decrease the resistivity to facilitate the direct integration of the equations of MHD. While quenching does decrease the resistivity at 0.1 mbar, the largest resistivities occur between 10 and 100 mbar, near the quench point of the ions, at sufficiently high pressures where the ions are not homogenized on the nightside. It is thus capturing the dynamics of the magnetic field on the nightside at these pressures that will hinder MHD modelling efforts.  Despite these challenges, true MHD models that solve for the evolution of the magnetic field are still required as we show that the magnetic Reynolds number, computed from both magnetic drag and drag-free simulations, can become significantly greater than unity on parts of the dayside, and when chemical kinetics is considered, parts of the nightside as well. These high Reynolds numbers indicate that the possibility exists to deform the field, violating an underlying assumption of the magnetic drag model that the field is static, thus warranting caution in over-interpreting the results. This possibility for field deformation only increases with temperature, as is highlighted by the MHD simulations of \citet{rogers_2014}, \citet{rogers_2014b}, and \citet{rogers_2017}, and we advocate for prudence in making quantitative determinations from magnetic drag models applied to ultra-hot Jupiters.

While we demonstrate here that chemical kinetics of the charged alkali species plays an important role in modelling magnetic effects in hot Jupiter atmospheres, we conclude by noting that this investigation is by no means comprehensive and a number of potentially important avenues remain unexplored. Perhaps most apparent is the potential for photoionisation to influence the ionisation of potassium on the dayside, especially around the substellar point, as highlighted in \citet{lavvas_2014}.  This increase in dayside ionisation could, through the transport demonstrated here, further increase the nightside ionisation and increasing the coupling to the magnetic field. Clouds and hazes may serve to attenuate any incoming photoionising radiation, limiting its importance; however, they may also serve as charge carriers, as in the atmosphere of Titan \citep{lavvas_2010}, and they may themselves become sources for electrons, as in protoplanetary disks \citep{desch_2015}.  Accounting for these effects due to aerosols could change not only the ionisation fraction but also the mass of the dominant charge carriers, if the aerosols became sufficiently charged, altering the conductivity of the atmosphere and thus any magnetic coupling. The charging of aerosols may also contribute to the occurrence of lightning within the atmosphere \citep{helling_2019b}.  While all of these should be considered in future modelling efforts, the most significant potential contribution to our understanding of magnetic effects remains the self-consistent modelling of the evolution of the magnetic field that accounts for any nightside decoupling of the field, as discussed here and elsewhere in the literature \citep[e.g.,][]{seidel_2026}. 

\section*{Acknowledgements}

We would like to thank the anonymous referee for their thoughtful comments which improved the quality of the manuscript. We would also like to thank Catherine Walsh for her help with finding references for reaction rates.  DAC is supported by the Max Planck Society. This research was also supported by a UK Research and Innovation (UKRI) Future Leaders Fellowship MR/T040866/1, and partly supported by the Leverhulme Trust through a research project grant RPG-2020-82 alongside a Science and Technology Facilities Council (STFC) Small Award ST/Y00261X/1. JEO is supported by a Royal Society University Research Fellowship.

The analysis of the simulation data made use of the following {\sc python} packages: {\sc aeolus} \citep{sergeev_2024}, {\sc iris} \citep{hattersley_2023}, {\sc matplotlib} \citep{hunter_2007}, {\sc numpy} \citep{harris_2020}, and {\sc scipy} \citep{virtanen_2020}.

\section*{Data Availability}

The simulation data are available for download from the Zenodo online repository at \href{https://doi.org/10.5281/zenodo.22274728}{doi.org/10.5281/zenodo.22274728}. For the purpose of open access, the authors have applied a Creative Commons Attribution (CC BY) licence to any Author Accepted Manuscript version arising.



\bibliographystyle{mnras}
\bibliography{references} 




\appendix

\section{Comparison with V19-Only Models}
\label{Appendix:V19Comp}
Previous models of HD~209458~b performed using the UM \citep[e.g.,][]{zamyatina_2023} used the \citetalias{venot_2019} network with the atomic alkali abundances determined from an approximate threshold method from \citet{amundsen_2016} wherein the atomic alkali species are included for $T > T_{\mathrm{crit},s}(P)$ for some critical temperature $T_{\mathrm{crit},s}(P)$ for species $s$. More recently, simulations were performed using the UM which employed a lookup table of equilibrium alkali abundances \citep{christie_2026}. In both cases, these alkali abundances do not impact the chemical network directly, and only influence the C/N/O/H chemistry indirectly through the impact on the opacities and thus the temperatures.

To test whether including alkali species within the chemical network has a significant impact on the non-alkali abundances, we run our fiducial non-magnetic HD~209458~b case, solving the chemical kinetics for only the C/N/O/H chemistry using the \citetalias{venot_2019} network, with the alkali abundances determined via equilibrium table lookup as in \citet{christie_2026}. We then compare the results to the fiducial non-magnetic runs described in the main text. The simulations were run for 500 days.

\begin{figure*}
	\includegraphics[]{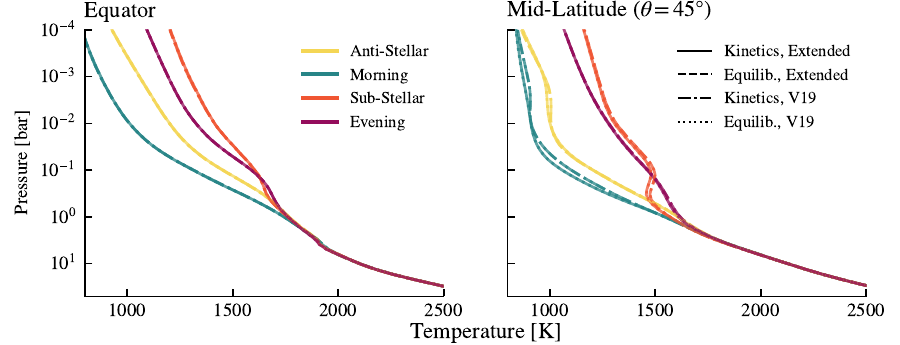}
    \caption{The equatorial (left) and mid-latitude (right) pressure-temperature profiles for the fiducial instellation cases without drag as well as chemical kinetics and equilibrium chemistry simulations using only the \citetalias{venot_2019} network, as described in the text. The lines for the two kinetics models overlap, as do the lines for the equilibrium models. }
    \label{Fig:pt_v19}
\end{figure*}

\begin{figure*}
	\includegraphics[]{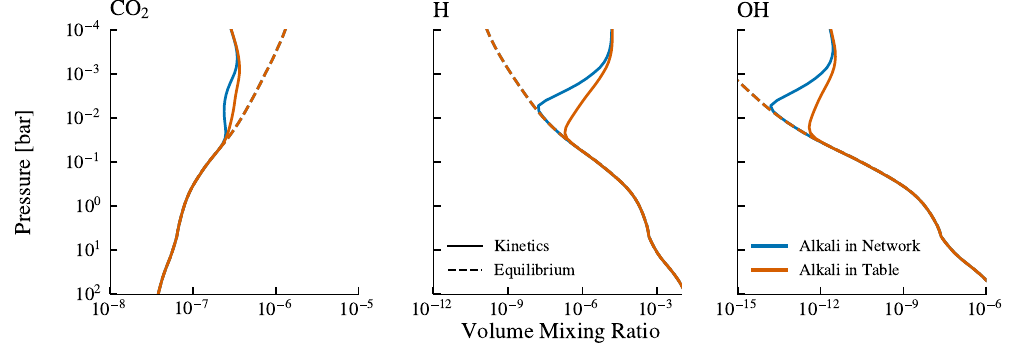}
    \caption{Equatorial abundances at the morning terminator of \ce{CO2} (left), \ce{H} (middle), and \ce{OH} (right). The inclusion of alkali species in the chemical network instead of including them only in an opacity lookup table leads to a small difference in the \ce{CO2} abundance on the morning terminator.}
    \label{Fig:chem_v19}
\end{figure*}

The atmospheric temperatures are not meaningfully impacted by the choice of alkali scheme, with the kinetics and equilibrium cases both differing by less than 5~K between simulations including alkali species in the chemical network compared to simulations employing a look-up table (see Figure \ref{Fig:pt_v19}). Similarly, we find the non-alkali abundances to be minimally impacted by the inclusion of the alkali species in the chemical network, with \ce{CH4}, as an example, differing by less than 3\% in the kinetics cases and less than 10\% in the equilibrium cases. The exception is \ce{CO2} where equilibrium abundances differ by less than 2\% but by up to 50\% at pressures between 1 and 10 mbar on the morning terminator in the kinetics case. The \ce{CO2} difference in the kinetics case is likely due to the impact of the alkali chemistry on the \ce{H} and \ce{OH} radical abundances, with these radicals differing by up to 500\% between kinetics simulations (see Figure \ref{Fig:chem_v19}). 

While the \ce{CO2} is likely impacted directly by the inclusion of alkali species in the chemical network, we note that some temperature differences may also arise via the implementation of the opacity tables. Opacity sources within the chemical network have their abundances updated on the chemical timescale ($\Delta t_\mathrm{chem}=3750\,\mathrm{s}$) while the abundances retrieved from the look-up tables are updated every call to the radiation solver ($\Delta t_\mathrm{rad}=150\,\mathrm{s}$)\footnote{This is also the case for the threshold approach to alkali species implemented in \citet{amundsen_2016}.}. This choice in implementation is due to the negligible computational overhead associated with the table. Although this may introduce differences in temperature, as the total differences are already small, as discussed above, we do not investigate this further.

We conclude that for the parameter space investigated here the chemical kinetics alkali species do not impact the non-alkali species, except for any impact via the inclusion of magnetic drag, as discussed in the main text. This may not be the case for hotter atmospheres, where ionisation and quenching can significantly deplete the total abundance of neutral atomic alkalis, thus altering the opacity. Similarly, quenching in cooler atmospheres may see increased atomic alkali abundances at low pressure due to quenching suppressing the chloride abundances, although this remains speculative. Both of these cases are beyond the scope of this paper.

\section{Radiative Recombination}
\label{Appendix:RR}
\begin{figure}
	\includegraphics[]{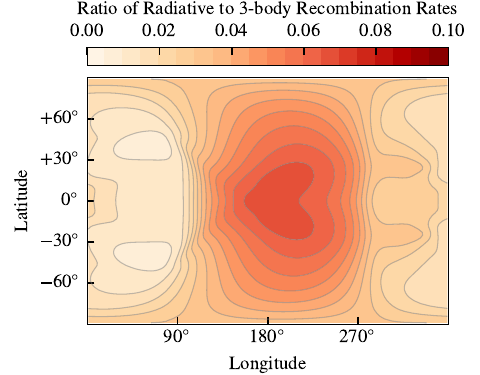}
    \caption{The ratio of the radiative recombination rate and the three-body recombination rate for potassium at a pressure of $3\times 10^{-5}$ bar for the non-magnetic equilibrium case with the fiducial instellation.}
    \label{Fig:recomb_ratio}
\end{figure}

In the model presented here, we do not consider radiative processes impacting the ion abundances, specifically, photoionisation and radiative recombination. While \citet{lavvas_2014} establish that photoionisation could impact the substellar ionisation fraction as deep as 10 mbar, the importance of radiative recombination (e.g., \ce{K+ + e- -> K + h$\nu$}) at pressures considered here is not as established. To understand how the radiative recombination rate \citep{verner_1996} ,

\begin{equation}
k_\mathrm{rr}=4.6423\times 10^{-10}T^{-0.841}\exp\left(-60/T\right)\,\, ,
\end{equation}

\noindent may alter the ionisation fraction, we examine the ratio of the radiative and 3-body recombination rates,

\begin{equation}
\frac{k_\mathrm{rr}x_\mathrm{e}x_\mathrm{K^+}n_\mathrm{tot}^2}{k_\mathrm{8}x_\mathrm{e}x_\mathrm{K^+}n_\mathrm{tot}^3} \,\, ,
\end{equation}

\noindent where $x_\mathrm{e}$ and $x_\mathrm{K^+}$ are the volume mixing ratios of electrons and potassium ions, and $n_\mathrm{tot}$ is the total particle number density. The ratio on an isobaric surface at a pressure of $3\times 10^{-5}$ bar is shown in Figure \ref{Fig:recomb_ratio}. This pressure represents the lowest pressure isobaric surface entirely within the computational domain and thus approaches the upper boundary on the dayside. On this surface we observe that at its largest, the radiative recombination rate is ten percent of the three-body recombination rate, and as the ratio scales as $n_\mathrm{tot}^{-1} \sim p^{-1}$, the radiative rate will become less important relative to three-body recombination deeper in the atmosphere, in the bulk of the computational domain. At 1 mbar, for example, the radiative rate peaks at only 0.3\% of the three-body recombination rate. We conclude that radiative recombination will not appreciably alter the ionisation fraction, although we note that were we to extend the model to higher altitudes and thus lower pressures, this would not be the case.

\section{Additional Plots}
\label{Appendix:AddPlots}

\begin{figure*}
	\includegraphics[]{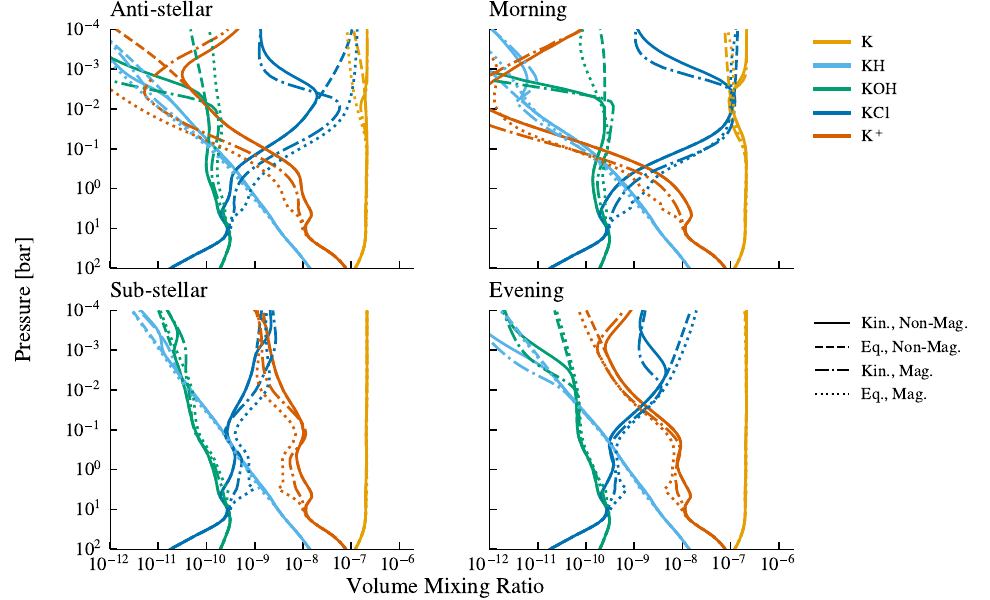}
    \caption{The equatorial abundances of K-bearing species for the $1.5\times$ the fiducial HD~209458~b instellation cases. }
    \label{Fig:chem_K_1.5x}
\end{figure*}

\begin{figure*}
	\includegraphics[]{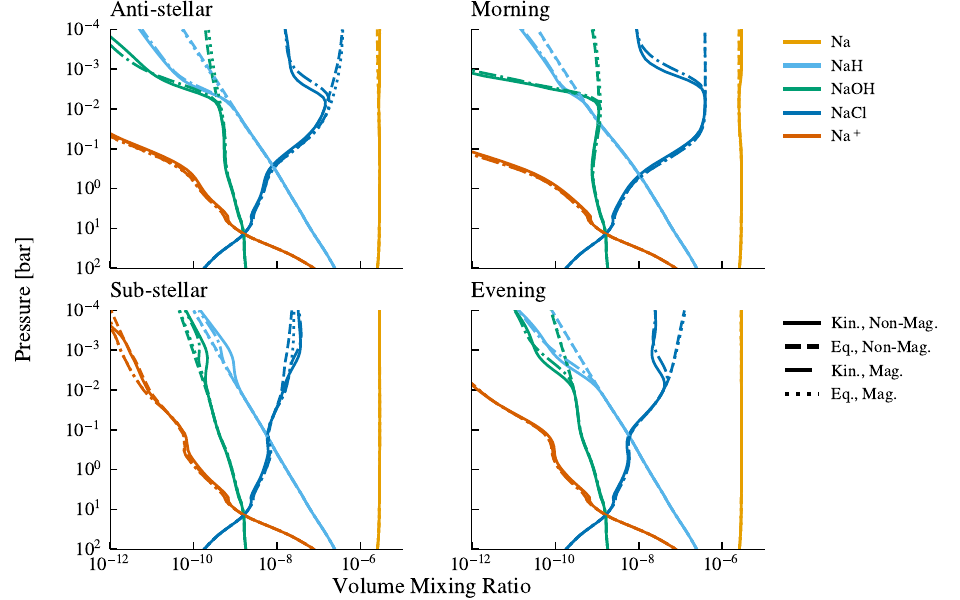}
    \caption{The equatorial abundances of Na-bearing species for the fiducial HD~209458~b instellation cases. Although departures from equilibrium values occur at pressures less than 1 mbar, the bulk of \ce{Na} remains in atomic form throughout the atmosphere.}
    \label{Fig:chem_Na_1x}
\end{figure*}

\begin{figure*}
	\includegraphics[]{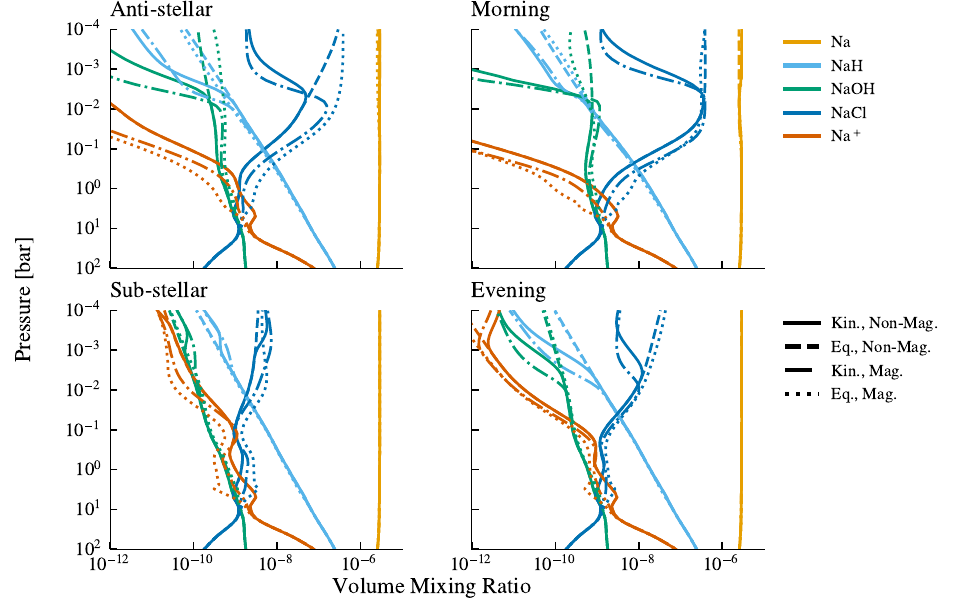}
    \caption{The equatorial abundances of Na-bearing species for the $1.5\times$ the fiducial HD~209458~b instellation cases.  As in the fiducial instellation case, the bulk of the \ce{Na} remains in atomic form throughout the atmosphere.}
    \label{Fig:chem_Na_1.5x}
\end{figure*}

\begin{figure*}
	\includegraphics[]{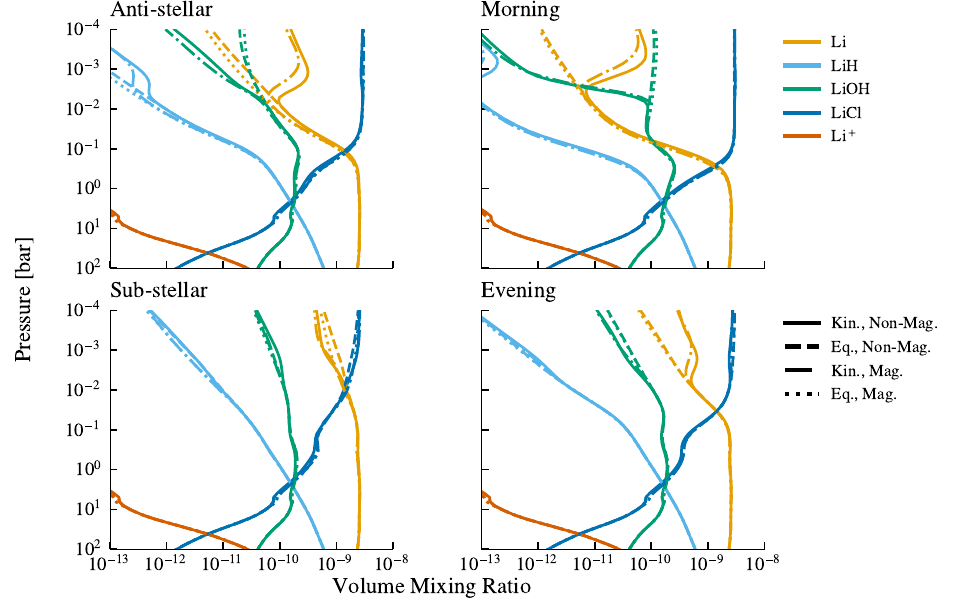}
    \caption{The equatorial abundances of Li-bearing species for the fiducial HD~209458~b instellation cases.  \ce{Li} transitions from being atomic in the deep atmosphere to being predominantly in the form of \ce{LiCl} in the upper atmosphere.  Chemical kinetics does increase the abundance of atomic \ce{Li} above $\sim 10$ mbar, but the volume mixing ratio remains less than $10^{-9}$.}
    \label{Fig:chem_Li_1x}
\end{figure*}

\begin{figure*}
	\includegraphics[]{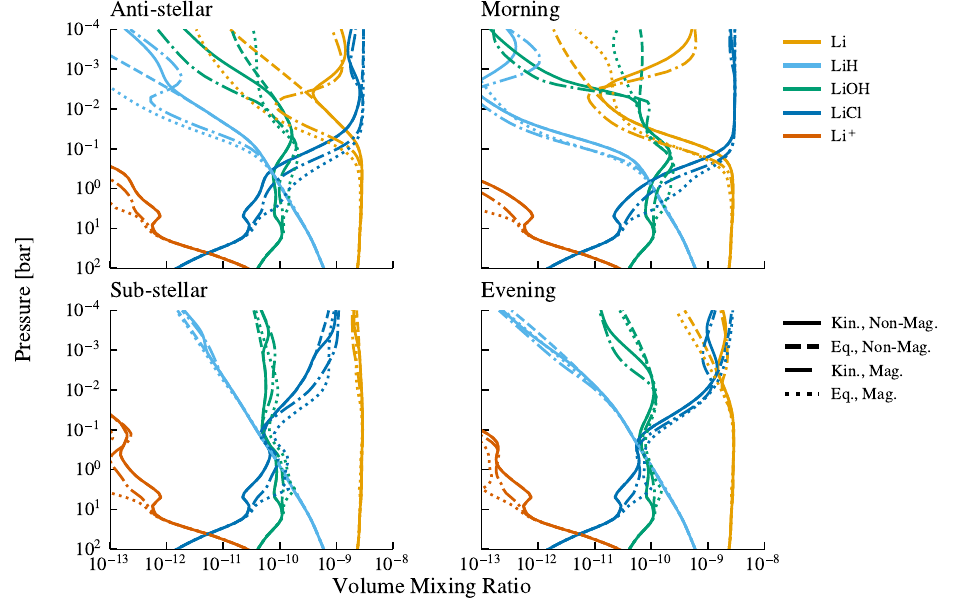}
    \caption{The equatorial abundances of Li-bearing species for the $1.5\times$ the fiducial HD~209458~b instellation cases. \ce{Li} is primarily atomic at the substellar point, and while it remains atomic in the deep atmosphere, at lower pressures it forms \ce{LiCl} as it is transported to the cooler nightside. }
    \label{Fig:chem_Li_1.5x}
\end{figure*}

In this appendix we present additional plots that might be of interest to the reader. Figures \ref{Fig:chem_K_1.5x} to \ref{Fig:chem_Li_1.5x} show equatorial abundances for K-, Na-, and Li-bearing species, supplementing Figure \ref{Fig:chem_K} in the main text.

\bsp	
\label{lastpage}
\end{document}